\documentclass[11pt,a4paper]{article}

\usepackage[a4paper,margin=1in]{geometry}
\usepackage{amsmath}
\usepackage{amssymb}
\usepackage{graphicx}
\usepackage{booktabs}
\usepackage{tabularx}
\usepackage{caption}
\usepackage{subcaption}
\usepackage{siunitx}
\usepackage[hyphens]{url}
\usepackage[square,numbers,sort&compress]{natbib}
\usepackage[hidelinks]{hyperref}
\usepackage{xspace}

\graphicspath{{figures_public/}}
\newcommand{\figwidth}{0.85\linewidth}
\newcommand{\cotwoeqsym}{\mathrm{CO_2\text{-}eq}}
\newcommand{\cotwoeq}{\ensuremath{\cotwoeqsym}\xspace}
\newcommand{\nparsym}{\bar{N}_{\mathrm{parallel}}}
\newcommand{\npar}{\ensuremath{\nparsym}\xspace}

\title{Greenpixie's AI Token Methodology\\[14pt]
\large Assessing the Energy, Water and \cotwoeq Impact of AI Tokens for Open and Closed Weight Models%
}
\author{Joshua Horswill, Ross Hunter, Matt Clifford, James Hall}
\date{Greenpixie Ltd., London, UK.\\[4pt]}

\begin{document}
\maketitle

\vspace{0.2em}
\begin{center}
\includegraphics[width=0.3\textwidth]{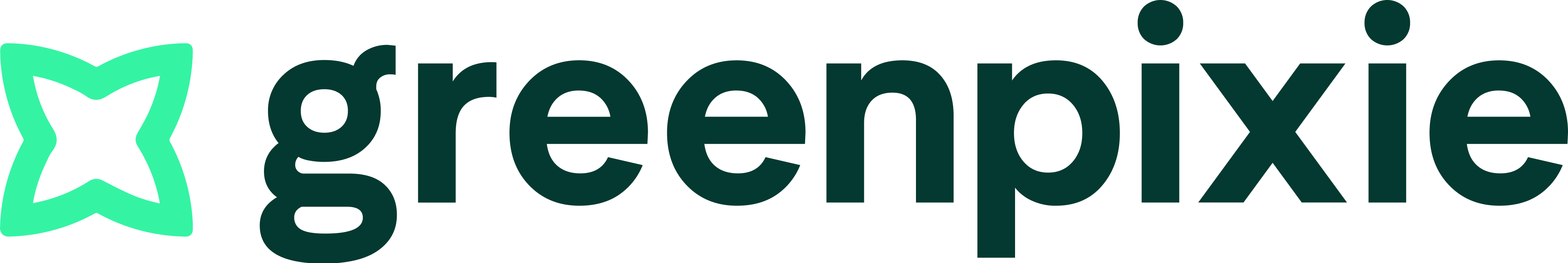}
\end{center}
\vspace{1em}

\begin{center}
\footnotesize
\copyright~2026 Greenpixie Ltd. This work is licensed under a Creative Commons
Attribution-NonCommercial-NoDerivatives 4.0 International licence (CC BY-NC-ND
4.0),\\ \url{https://creativecommons.org/licenses/by-nc-nd/4.0/}. Commercial use is not permitted.
\end{center}

\begin{abstract}
We describe a methodology for estimating the per-token energy cost of
cloud-hosted large language model (LLM) inference, separating between
input (prefill) and output (decode) tokens.
Graphics processing unit (GPU) energy usage is measured during inference benchmarking with open-weights models on a wide range of text-based tasks.
The remaining server energy contribution from non-GPU hardware
is estimated from the inference wall time. Bayesian linear regression is used to model the relationship between energy per token and LLM size,
request traffic, and hardware deployment configuration.
Proprietary frontier LLMs of unknown size and deployment are binned into size
buckets based on naming conventions and performance priors, and the space of possible LLM configurations is sampled with Monte-Carlo methods to
give a representative average energy per token and uncertainty.
We also describe how these energy measurements can be used to estimate the carbon-dioxide equivalent (\cotwoeq) emissions, both usage and embodied,
and water consumed per token of AI inference.
This methodology provides actionable data that enables reductions in cost, electricity usage, \cotwoeq emitted and water consumed in cloud and Software as a Service (SaaS).
\end{abstract}

\newpage

\section{Introduction}
\label{sec:intro}

The environmental impact and financial cost of AI has become one of the most discussed topics in the fields of GreenOps and FinOps~\citep{linuxfoundation_tokenomics_2026}.
Although the current cloud spend on AI is fractional compared to traditional cloud services,
this relationship will change dramatically in the near future. Goldman Sachs Research
forecasts consumption to multiply $24\times$ between 2026-2030
to $120$ quadrillion tokens\footnote{LLMs split words into short character sequences called tokens, which are then mapped to numerical representations that the model can process. This is done by a tokenizer. Tokenizers differ, so the same input text will map to different numbers of tokens for different LLMs~\citep{trott2024tokenization}.} per month~\citep{goldman_token_growth_2026}.

This enormous growth in usage necessitates enormous growth in AI infrastructure
and the electricity needed to power it.
The global electricity consumption of data centres is projected to more
than double between 2024 and 2030, from $\sim 415$~TWh to $\sim 945$~TWh~\cite{iea_energy_and_ai_2025}.
This figure is comparable to the current annual electricity
consumption of Japan~\citep{iea_japan_electricity}. The main driver is
accelerated servers for AI,
growing at $\sim 30\%$ per year, against roughly $9\%$ per year for conventional
servers~\citep{iea_energy_and_ai_2025}. A United Nations University
study projects that AI-related water consumption could equal the
basic annual domestic needs of $1.3$ billion people by the end of the
decade, with AI infrastructure generating up to $2.5$ million tonnes
of e-waste annually by 2030~\citep{un_ai_environment_2026}.

To help mitigate the environmental impact of this growth, it is crucial that enterprises have
robust and actionable sustainability data
on their AI usage as it scales. This will enable decision-making based
on both cost (FinOps) and sustainability (GreenOps) grounds. This paper details
Greenpixie's methodology for estimating the per-token cost in electricity used, \cotwoeq emissions generated
and water consumed due to large language model (LLM) inference in the public cloud.
The methodology's primary aim is to provide useful and defensible estimates for
enterprise use.

The methodology is based on measuring energy per input (prefill) and output (decode) token
for a number of open-weights LLMs under different workloads and configurations.
The results are then mapped to the major frontier proprietary models with a quantified
uncertainty to enhance actionability.
This is believed to be, at the time of writing, the only methodology that
combines realistic quantisations and batching on current hyperscaler hardware across
1.8B to 1T parameters with a prefill/decode split.
The latter split is vital for two reasons. Firstly, input and output tokens are usually
billed separately at different prices by the LLM providers. Secondly, input tokens
are processed in parallel during the prefill phase, which is compute-bound. During
decode, output tokens are emitted one at a time (autoregressively), which is
memory-bandwidth-bound~\citep{llm_inference_handbook_prefill_decode}. We find that
these two inference phases have drastically different energy per token profiles.
To reduce cost and energy per token, realistic deployments also interleave requests
from different users and process them in parallel in a single batch, with continuous
admission and retirement of parallel requests as conversations end and compute capacity
is freed~\citep{llm_inference_handbook_batching}. In order to cleanly
attribute energy to input and output tokens, we approximate this by dispatching
all requests in a batch at the same time and allow all requests to complete before
dispatching the next batch. We test LLMs with classical ``dense''
architectures, in which every parameter participates in each forward pass, and
mixture-of-experts (MoE) models, in which a routing network
selects a task-appropriate subset of the parameters per token.
Understanding the consequences of these details for energy is crucial for
accurately covering the plethora of enterprise AI use cases and hyperscaler
AI products.

The remainder of this paper is organised as follows.
Section~\ref{sec:related} reviews related work on AI-inference energy
estimation. Section~\ref{sec:methodology} describes the benchmarking
configuration, the energy attribution and calculation, the regression
model and the conversion of energy to \cotwoeq and water.
Section~\ref{sec:results} shows the regressions on energy and wall clock time per token.

Section~\ref{sec:mapping} maps these open-weights regression fits to proprietary
models. Section~\ref{sec:perf_emissions_tradeoff} discusses the
decision levers the numbers support and sets out the limitations
and future work. A conclusion is given in Section~\ref{sec:conclusion_summary}.

\section{Related work}
\label{sec:related}

A variety of other studies have assessed the environmental impact of AI usage.
Jegham et al.~\citep{jegham2025hungryaibenchmarkingenergy} use API token throughput
data to infer LLM deployment configurations, which are combined with assumptions on the server utilisation, batching configuration and proprietary model size
to give an estimate of the per-token energy cost of inference from the major providers.
However, the impact on energy per token of many of these assumptions
is not fully quantified, and queuing and networking effects could
bias energy measurements reliant on API throughput, in particular for memory-bandwidth-bound decoding. Epoch
AI~\citep{epoch_chatgpt_energy_2025} and Oviedo et al.~\citep{oviedo2026energy} have provided
bottom-up arithmetic and Monte-Carlo based estimates of per-query energy on
H100 deployments, but did not split the per-query energy between input and output tokens.
Vartziotis et al.~\citep{vartziotis2026tokens_to_wh} provide a thorough calibrated analytical
model of the GPU energy cost of LLM inference, but do not calibrate it with
their own experimental data, nor extend it to cover non-GPU energy costs.

Luccioni et al.'s life-cycle assessment of BLOOM~\citep{luccioni2023bloom}
directly measured the energy costs of training and serving an API-deployed
LLM, but did so without batching on a now old architecture. The AI
Energy Score~\citep{ai_energy_score_2024}, building on the
benchmarking of Luccioni et al.'s ``Power Hungry
Processing''~\citep{luccioni2024power_hungry}, ranks open-weights
models by GPU energy per $1{,}000$ queries per task, with one-to-five-star efficiency ratings.
The ML.ENERGY project~\citep{chung2025mlenergy_benchmark, chung2026joules}, Wilkins et al.~\citep{wilkins2024offline_energy} and Vellaisamy et al.~\citep{vellaisamy2026token_energy}
ran their own inference benchmarks of open-weights LLMs,
measuring energy directly from the GPU telemetry, and TokenPowerBench~\citep{niu2026tokenpowerbench} additionally publish
their framework on GitHub for running such benchmarks. The latter two studies also perform a prefill/decode split,
but none of them extend coverage to frontier proprietary LLMs.

Stojkovic et al.~\citep{stojkovic2024llmenergy} report that
per-token inference energy scales exponentially with
prompt and response length over the benchmarked range.
Fernandez et al.~\citep{batchsize_energy_acl2025}
suggest that it decays exponentially with batch size.
These studies motivate our measurement of per-token energy efficiency as a function of the number of parallel requests \npar, and motivates analogous sequence-length measurements in the future.
Further measurement frameworks and academic studies exist~\citep{henderson2020systematic,wu2022sustainable,samsi2023wordstowatts,desislavov2023inference}, but leave space
for applying a methodology in a business context.

EcoLogits~\citep{ecologits} provide a hybrid approach, where ML.ENERGY
energy measurements on open-weights models are combined with proprietary
and open-weights token throughputs from OpenRouter.
CodeCarbon~\citep{lottick2019codecarbon} instruments a running job
rather than predicting from public information, and integrates
EcoLogits for LLM inference, providing a route to using those estimates
at scale and commercially, but not in the cloud.

Google's 2025 disclosure of
the per-token energy, \cotwoeq, and water footprint of Gemini-scale
deployment~\citep{google_gemini_emissions_2025} is the closest
hyperscaler-internal counterpart to this work, and Mistral has published a third-party-reviewed life-cycle assessment for Mistral
Large~2~\citep{mistral_large_2_lca}, covering training
and inference. Some headline figures exist with no accompanying
methodology, such as Sam Altman's per-query ChatGPT energy and water
numbers~\citep{altman2025gentlesingularity}, but these offer no basis
for scrutiny or reproduction, and have not been updated since mid 2025.

To the best of our knowledge, no single study covers open-weights
models from 1.8B up to 1T total parameters on current hyperscaler
hardware at production quantisations, differentiating prefill and
decode tokens, and none applies the resulting predictions directly to
cloud-service stock keeping units (SKUs) and SaaS products without
further customer-side modelling. Prior studies have also not
sufficiently quantified the uncertainty associated with their assumptions.
We quantify the statistical and systematic uncertainties
in our methodology and provide an estimation range, so that our
method can be compared against others and provide best/worst case
scenarios when forecasting.

\section{Benchmarking methodology}
\label{sec:methodology}

Energy per token measurements are conducted by sampling the GPU power draw and inference time of deployed LLMs
as they process and answer a series of standardised prompts.

Instantaneous GPU power draw is measured with the NVIDIA Management
Library (NVML)~\cite{nvidia_nvml} and timestamps of the start and end of the different inference phases
are provided by the vLLM inference engine~\cite{vllm_kwon2023}.
We study and determine the relationships between energy per token and LLM size; LLM quantisation; LLM active parameter
count; hardware type and quantity; and \npar. With these fitted relationships,
we map to proprietary LLMs and quantify the uncertainty in that mapping.

The non-GPU energy per token is not measured directly, but estimated using server power draw
figures from Greenpixie's compute-energy methodology and multiplying the power by the measured inference time per token.
An energy uplift is applied to account for the estimated cost of training the LLM.
The energy usage is also translated into \cotwoeq emissions and water consumed
to give a full account of the environmental impact.

\subsection{Benchmarking configuration}
\label{sec:setup}
Benchmarks are run on NVIDIA H100 PCIe (80~\si{\giga\byte}
HBM2e)~\citep{nvidia_h100_pcie} and NVIDIA B200 (192~\si{\giga\byte}
HBM3e)~\citep{nvidia_b200} cards, rented through Runpod~\citep{runpod_docs},
with vLLM 0.14.0 (Python) \citep{vllm_kwon2023} on top of
PyTorch~2.8.0, CUDA~12.8, cuDNN~9 and Python~3.11.
The vLLM configuration is shown in Table~\ref{tab:vllm_settings}.
Maximum \npar count\footnote{
The average number of requests processed in parallel can sit well below the configured
maximum. Stragglers that do not fit into available GPU memory are queued until
VRAM is freed, depending on the memory available, the
average prompt length, and the LLM's attention-head architecture.
} (vLLM's \texttt{max\_num\_seqs}) is swept across
runs from~$16$ to~$2048$. GPU board power is sampled
every $\Delta t = 10~\si{\milli\second}$. Tensor parallelism is set
automatically to the detected GPU count.

The full list of benchmarked LLMs is given in Appendix~\ref{app:model_tables}.
Models are quantised to FP8 and NVFP4 precision.
Their workload covers text-inference tasks (instruction following,
reasoning, knowledge recall, code completion, classification, and
summarisation) drawn from the following datasets:
IFEval~\citep{zhou2023ifeval},
Alpaca~\citep{taori2023alpaca},
BBH~\citep{suzgun2022bbh},
GSM8K~\citep{cobbe2021gsm8k},
ARC-Challenge~\citep{clark2018arc},
HLE~\citep{phan2025hle},
GPQA~\citep{rein2023gpqa},
MMLU-Pro~\citep{wang2024mmlupro},
HumanEval~\citep{chen2021humaneval},
SQuAD-v2~\citep{rajpurkar2018squadv2},
CNN-DailyMail~\citep{hermann2015cnndaily},
XSum~\citep{narayan2018xsum},
IMDB~\citep{maas2011imdb}, and
WikiText~\citep{merity2016wikitext}. Each model is requested to perform each
task in each dataset, giving prompt and response lengths up to roughly
$14{,}000$ input tokens and $8{,}192$ output tokens per request.
Note that, whilst frontier models may saturate these benchmarks in terms of accuracy,
this does not directly affect the energy per token the model uses.

\begin{table}[htbp]
  \centering
  \caption{vLLM engine settings held constant across the campaign.}
  \label{tab:vllm_settings}
  \small
  \begin{tabularx}{\linewidth}{l l X}
    \toprule
    \textbf{Setting} & \textbf{Value} & \textbf{Role} \\
    \midrule
    \texttt{max\_num\_batched\_tokens} & 65{,}536 & Cap on the total number of tokens in any single batched forward pass. \\
    \texttt{enable\_chunked\_prefill}  & True      & Splits long prefills into chunks so they dynamically interleave with active decodes, matching production serving. \\
    \texttt{max\_tokens}               & 8{,}192   & Maximum tokens generated per request. \\
    \texttt{gpu\_memory\_utilization}  & 0.90      & Fraction of GPU memory allocated to the KV cache + weights. \\
    \texttt{distributed\_executor\_backend} & \texttt{"mp"} & Multi-process executor (rather than Ray). \\
    \texttt{pipeline\_parallel\_size}  & 1         & No pipeline parallelism; tensor parallelism set to detected GPU count. \\
    \bottomrule
  \end{tabularx}
\end{table}

\subsection{Energy attribution}
\label{sec:attribution}

Energy attribution is request-scoped: a user
is responsible only for the runtime that their request causes, not
for idle power between requests. When
multiple users' requests share a forward pass, the runtime of that pass
is divided fractionally between the participating requests in
proportion to the request lifetime. Idle power consumed by an underutilised
inference server between requests is not attributed to any user.

GPU board power $P_i$ is sampled at uniform intervals~$\Delta t$, producing a
time series $(t_i, P_i)$, which is linearly interpolated to any time $s$
between consecutive samples as

\begin{equation}
  \hat{P}(s) = P_i + \frac{s - t_i}{t_{i+1} - t_i}\,(P_{i+1} - P_i),
  \qquad t_i \le s \le t_{i+1}.
\end{equation}

Each sample interval that overlaps an inference phase~$\phi$ is clipped to
that phase, $s_i^{(b)} = \max(t_i^{(b)},\; t^{\phi}_{\mathrm{start}})$ and
$s_{i+1}^{(b)} = \min(t_{i+1}^{(b)},\; t^{\phi}_{\mathrm{end}})$, so that
power samples straddling a phase boundary do not leak between phases.
For an inference phase~$\phi \in \{$ prefill-only,
overlap between prefill and decode\footnote{When continuously-batching requests like this,
chunks of one request’s prefill will run in the same forward
pass as other requests’ decode.},
decode-only $\}$,
the linearly-interpolated power draw is integrated
via the trapezoidal rule to give the phase's energy as

\begin{equation}
  E_{\phi}
  \;=\; \sum_{b=1}^{B}
        \sum_{i=1}^{N_b - 1}
        \frac{\hat{P}(s_i^{(b)}) + \hat{P}(s_{i+1}^{(b)})}{2}
        \;\big(s_{i+1}^{(b)} - s_i^{(b)}\big),
  \label{eq:energy_integration}
\end{equation}

where the outer sum is over the $B$ batches that span phase~$\phi$ and the
inner sum is over the $N_b$ timestamp-power samples $t_i^{(b)}$ within batch~$b$
clipped to the phase boundaries $t^\phi_{\mathrm{start}}, t^\phi_{\mathrm{end}}$.
vLLM provides an endpoint at runtime that reports the start and end timestamps
of prefill and decode for each request.
The per-token energy is measured as $E_{\phi}$ summed across all batches,
divided by the total (input or output) tokens processed during phase
$\phi$ across all batches. Splitting inference
phases allows differentiation between input and output token energy.

Non-GPU contributions to the total energy consumption (vCPUs, RAM,
networking interface controller (NIC), fans, and power supply
unit (PSU) etc.) are estimated as the inference-phase wall
time multiplied by the power rating of the host server. AWS
\texttt{p5.48xlarge} ($8\times$H100) and \texttt{p6-b200.48xlarge}
($8\times$B200) nodes are used as reference server specifications.
Based on the AWS instance documentation~\citep{aws_accelerated_computing},
the chipsets for \texttt{p5} and \texttt{p6-b200} nodes are identified
as the Intel Xeon Sapphire Rapids (4th generation) and the Emerald Rapids
(5th generation) respectively. The default power assumes $50\,\%$ vCPU
utilisation, with RAM, fans, and PSU draw absorbed into the power curve.
For simplicity, the non-GPU power of these servers is split evenly across
each GPU, as is shown as a function of GPU count in
Table~\ref{tab:ec2_equivalent}. For comparison, a single H100 GPU draws around $350$~\si{\watt},
with a B200 taking up to $1$~\si{\kilo\watt}.

\begin{table}[htbp]
  \centering
  \caption{Assumed AI server specification used to estimate the
  non-GPU server contribution.}
  \label{tab:ec2_equivalent}
  \small
  \begin{tabular}{l r r r r}
    \toprule
    \textbf{GPU count} & \textbf{vCPU} & \textbf{Memory (GiB)} & \textbf{\texttt{p5} power (\si{\watt})} & \textbf{\texttt{p6-b200} power (\si{\watt})} \\
    \midrule
    1  &  24 &  256 &   35.14 &   60.81 \\
    2  &  48 &  512 &   70.28 &  121.62 \\
    4  &  96 & 1024 &  140.56 &  243.23 \\
    8  & 192 & 2048 &  281.13 &  486.46 \\
    16 & 384 & 4096 &  562.25 &  972.93 \\
    32 & 768 & 8192 & 1124.50 & 1945.85 \\
    64 & 1536 & 16384 & 2249.00 & 3891.71 \\
    128 & 3072 & 32768 & 4498.00 & 7783.42 \\
    \bottomrule
  \end{tabular}
\end{table}

\subsubsection{Training-energy uplift}
\label{sec:training_uplift}

Significant energy and emissions are incurred when training LLMs, which is
amortised over the model's lifespan by applying an uplift factor to the per-token
energy costs. The factor is the ratio of the total energy cost (training + inference)
divided by the total inference energy cost over the model's lifespan.
Multiple sources estimate that around 80-90\% of total energy costs is inference
for widely-used models~\citep{patterson2023sciencedirect, aws_inferentia_2019, nvidia_hpcwire_2019, mit_tech_review_2025}.
More lightly-served models may have a lower fraction, so we consider a range of
uplifts between $1.1\times$ (90\% inference; widely-served) to $1.67\times$ (60\% inference; lightly-served).
The range of uplift factors is propagated into the systematic uncertainty on the operational energy.

\subsection{Regression model fit to LLM cluster centroids}
\label{sec:regression}

In order to map our benchmarked energy per token values to different workload configurations
and proprietary LLMs, we fit a Bayesian linear regression model to the experimental results:

\begin{equation}
  \hat{y}
  \;=\; \hat{\beta}_0
       + \hat{\beta}_1\,N_{\mathrm{active}}
       + \hat{\beta}_2\,\mathrm{VRAM}_{\mathrm{GB}},
  \qquad
  \hat{\beta}_0,\hat{\beta}_1,\hat{\beta}_2 \geq 0,
  \label{eq:regression_centroid}
\end{equation}

where $N_{\mathrm{active}}$ is the active parameter count in billions
and $\mathrm{VRAM}_{\mathrm{GB}}$ is the total VRAM occupied by the LLM weights, in
gigabytes. Further architectural features are not considered, as they
would have to be approximated for proprietary LLMs, adding more uncertainty
rather than less.
$\hat{y}$ is one of four predicted target metrics: kWh per $1000$ (input,
output) tokens; and prefill (decode) wall clock time per $1000$ input (output) tokens.
Each fitted data point $\bar{y}_i$ is the mean of a cluster of
correlated experimental measurements (all tasks in all benchmark datasets etc.)
for the $i$-th LLM. Clusters contain around 80 individual
measurements on average. This clustering allows us to assume that the
uncertainties in each data point are approximately
independent.\footnote{In principle some correlation
exists across LLMs due to architectural and hardware variations.
Any underestimation of uncertainties due to this should be small relative
to the uncorrelated uncertainty.}
The 95\,\% credible interval on $\hat{\boldsymbol{\beta}}$ and the
centroid-level $R^2$ are therefore accurate.

The exact regression model is a conjugate Normal--Inverse--Gamma linear
regression under a vague prior, whose closed-form posterior separates
the \textit{epistemic} uncertainty associated with the position of the
regression line (which reduces with more benchmarked LLMs) from the
\textit{aleatoric} uncertainty related to the irreducible random
noise of the dataset.

We apply a non-negativity constraint by truncating the posterior: $\boldsymbol{\beta} \geq 0$.
This reflects the hypothesis that more active parameters
(more floating point operations per forward pass) and
larger memory footprints (more total VRAM footprint requires
more GPUs at fixed per-GPU memory, drawing more power) cannot
reduce per-token energy and wall clock time. Cluster centroids
more than five standard deviations from the regression line,
and individual energy and inference time per token measurements
more than three standard deviations from the cluster mean, are
excluded and the regression model is refit on the remaining clusters.

As noted by Refs.~\citep{stojkovic2024llmenergy, batchsize_energy_acl2025}, per-token energy depends non-linearly on \npar.
Rather than fit a single regression model, we
partition the data into subsets by hardware, LLM quantisation and \npar threshold, and fit
separately within each subset. For the proprietary-model predictions this
yields six subsets/regressions for output tokens (two
\npar regimes $\times$ three
hardware/quantisation configurations) and three for input tokens
(hardware/quantisation only, since per-input-token energy shows no
consistent \npar scaling), per target metric.
Low and high thresholds for the average \npar
axis are derived empirically to give a separation of at
least~$2\times$ in the target metric between the low and high bins.
This is inspired by the common $2\times$ price difference between batched and non-batched AWS SKUs for the same LLM.

\subsection{Statistical uncertainty quantification}
\label{sec:uncertainty_summary}

The energy efficiency per token for an LLM varies due to \npar, prompt and
response length, and any uncontrolled variables such as GPU temperature or
inference engine overhead. These variations represent uncertainty in the
predictions of the fitted regression models. We quantify this uncertainty
in terms of a 95\,\% credible band on the population line:

\begin{equation}
      \label{eq:cred-band}
\mathrm{CB}(\mathbf{x})
  \;=\; \mathbf{x}^{\!\top}\mathbf{m}_n \;\pm\;
        t_{0.975,\,2a_n} \cdot
        \sqrt{\,\tfrac{b_n}{a_n}\,
              \mathbf{x}^{\!\top} V_n\,\mathbf{x}\,},
\end{equation}

where $\mathbf{x} = (1,\; N_{\mathrm{active}},\;
\mathrm{VRAM}_{\mathrm{GB}})^{\!\top}$ is the query vector and
$(\mathbf{m}_n, V_n, a_n, b_n)$ are the Normal--Inverse-Gamma
posterior parameters of the centroid regression, where $n$ marks
quantities updated on all $n$ LLM centroids in the data ($n=0$ is the prior).
The regression model can be written in terms of the query vector, the coefficient vector $\boldsymbol{\beta}$, and the normal residual $\varepsilon$: $\bar{y}
= \mathbf{x}^{\!\top}\boldsymbol{\beta} + \varepsilon$, where
$\varepsilon \sim \mathcal{N}(0, \sigma^2)$, and $\sigma^2$ is the aleatoric variance on the residuals of the regression line $\bar{y}\ -\ \hat{y}$.
Since the prior $(\mathbf{m}_0, V_0, a_0, b_0)$ is vague, the posterior is data-driven only:
$\mathbf{m}_n$ coincides with the least-squares coefficients, $2a_n$ is the number of centroids
in the fit, and $b_n/a_n$ is the mean squared residual, which estimates $\sigma^2$.
The Student-$t$ multiplier with $2a_n$ degrees of freedom accounts for $\sigma^2$ being estimated
rather than known, so under the vague prior the band coincides numerically with the classical
95\,\% confidence band on the regression line.
Since $V_n$ is proportional to the covariance of $\boldsymbol{\beta}$, if the number and spread of centroids in a given direction of
feature space increases, the inner product $\mathbf{x}^{\!\top} V_n \mathbf{x}$ in this direction will decrease, and so will the epistemic uncertainty.
The aleatoric variance $\sigma^2$ remains the same size regardless; it is not part of the band on the line
and is added separately when predicting a single deployment in Section~\ref{sec:montecarlo}.

\subsection{Inference emissions and water consumption}
\label{sec:emissions_water}

Operational \cotwoeq emissions per token, denoted
$\cotwoeqsym_{\mathrm{op}}$ hereafter, are derived as the product of the IT-load inference energy
$E_{\mathrm{inference}}$, the provider-published power-usage effectiveness (PUE) for the cloud
region (accounting for cooling, lighting and power distribution at the data centre),
and the grid \cotwoeq intensity (GCI, in g\cotwoeq/kWh) for the cloud region at the particular time.
The operational emissions are then scaled up by $1.01\times$ to approximate on-site Greenhouse Gas
(GHG) Protocol Scope 1 emissions (backup generator fuel and refrigerant leakage)~\citep{ghg_protocol_corporate_standard}
at data centers, following Schneider Electric's estimate
that Scope 1 emissions for a representative data centre are 0.3--1.6\% of Scope 2
emissions~\citep{schneider_electric_scope3_2023}.

Greenpixie's proprietary methodology utilises region-specific PUE values, which
are taken from each provider's published sustainability reports where available
\citep{aws_sustainability, google_datacenter_efficiency,
azure_datacenter_sustainability}, with a global average applied as a
fallback. Hourly GCI is produced from a tiered hierarchy of online sources \citep{eia_open_data, epa_egrid, entsoe_transparency, uk_carbon_intensity_api, rte_eco2mix, ember_energy, ipcc_ar6_wg3_ch6, iea_emission_factors_2024}.
Emissions figures are produced in
accordance with the GHG Protocol Scope~2 Guidance~\citep{ghg_protocol_scope2}.

Water consumption is calculated by summing the direct cooling water
consumed by the data centre $W_{\mathrm{cool}} = \mathrm{WUE}_{\mathrm{cool}} \cdot E_{\mathrm{inference}}$
and the water consumed at the upstream power
station $W_{\mathrm{grid}} = E_{\mathrm{inference}} \cdot \mathrm{PUE} \cdot \mathrm{WCI}_{\mathrm{region}}$.
$\mathrm{WUE}_{\mathrm{cool}}$ is the published
provider-region water-usage effectiveness (WUE; litres of cooling
water per kWh of IT load), and $\mathrm{WCI}_{\mathrm{region}}$ is the World Resources
Institute regional water-consumption-intensity factor~\citep{wri_water_electricity_2020}.
This framework is informed by Ref.~\citep{li2023making_ai_less_thirsty}.

For some cloud AI line items the deployment region is not declared,
in which case global averages for the above coefficients are used.
Where the region is declared, it is possible that some compute
happens in other regions. Some serving stacks split prefill
and decode across separate compute pools, possibly
running in different data centres. Where this routing is not surfaced,
the methodology treats requests to the declared region at face value
in terms of grid \cotwoeq intensity and water-consumption
intensity.

Per-token cradle-to-gate embodied emissions, denoted
$\cotwoeqsym_{\mathrm{embodied}}$, are computed by amortising the
manufacturing emissions of the assumed hardware over
its operational lifetime (assumed six years), scaled by the inference wall-clock time
attributable to each request. All component formulas follow the
Boavizta methodology~\citep{boavizta_server, boavizta_api_manufacture}.

\section{Results}
\label{sec:results}

\begin{figure}[ht]
  \centering
  \includegraphics[width=\figwidth]{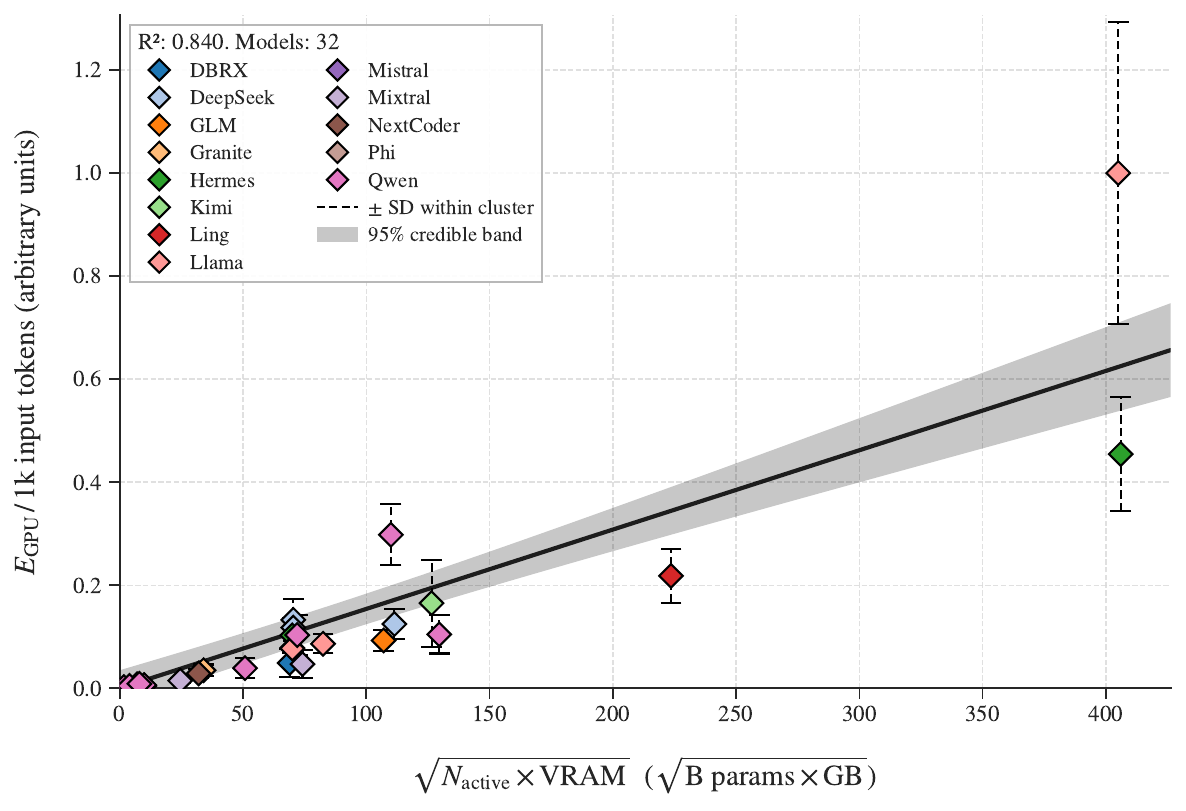}
  \caption{GPU energy usage per 1000 input tokens for all GPUs, quantisations and values of \npar, multiplied by a constant to mask the absolute scale.}
  \label{fig:headline_kwh_input}
\end{figure}

\begin{figure}[ht]
  \centering
  \includegraphics[width=\figwidth]{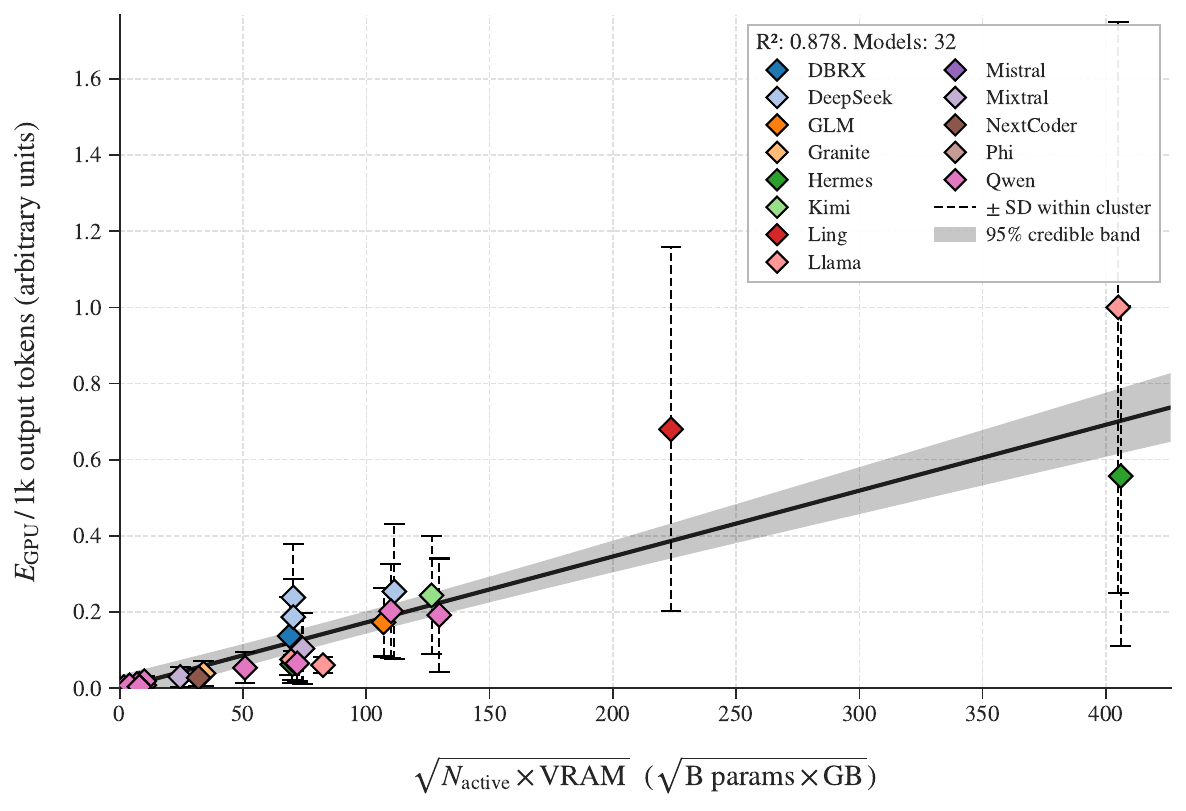}
  \caption{GPU energy usage per 1000 output tokens for all GPUs, quantisations and values of \npar, multiplied by a constant to mask the absolute scale.}
  \label{fig:headline_kwh_output}
\end{figure}

\begin{figure}[ht]
  \centering
  \includegraphics[width=\figwidth]{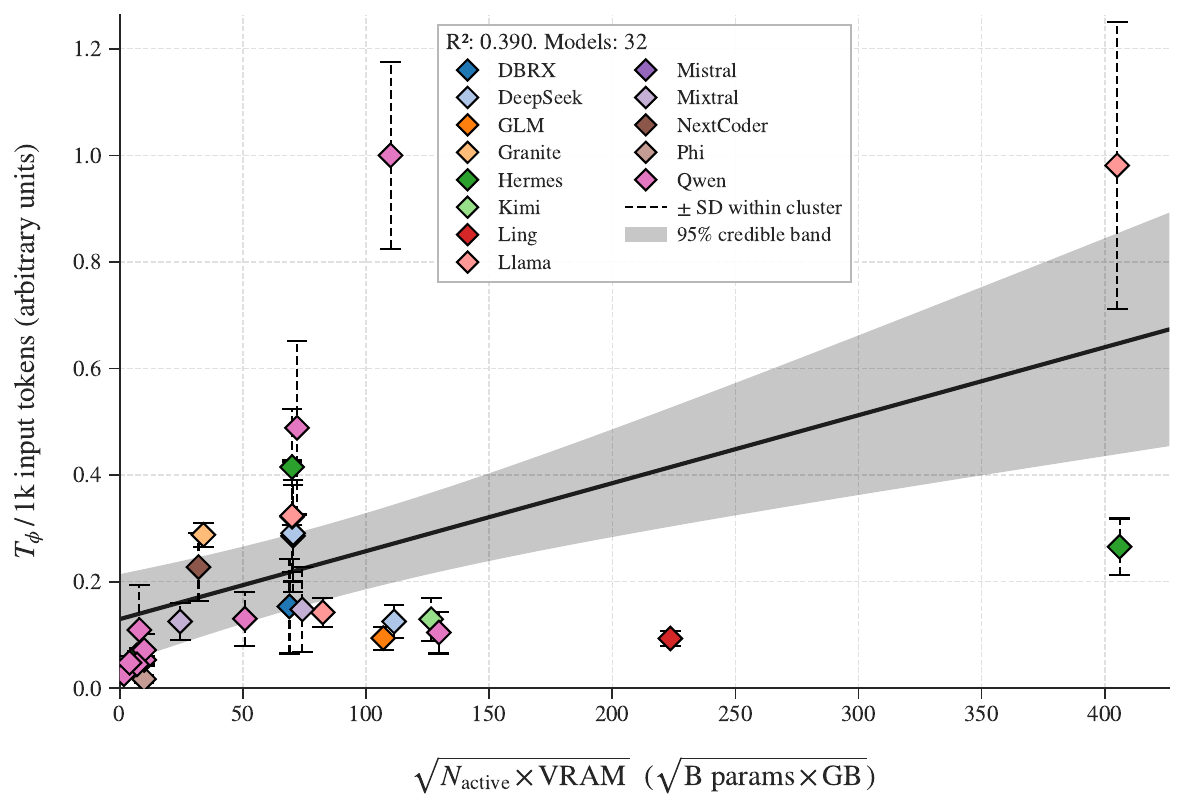}
  \caption{Prefill-phase per-token wall clock time $T_{\phi}$ for all GPUs, quantisations and values of \npar, multiplied by a constant to mask the absolute scale.}
  \label{fig:walltime_prefill}
\end{figure}

\begin{figure}[ht]
  \centering
  \includegraphics[width=\figwidth]{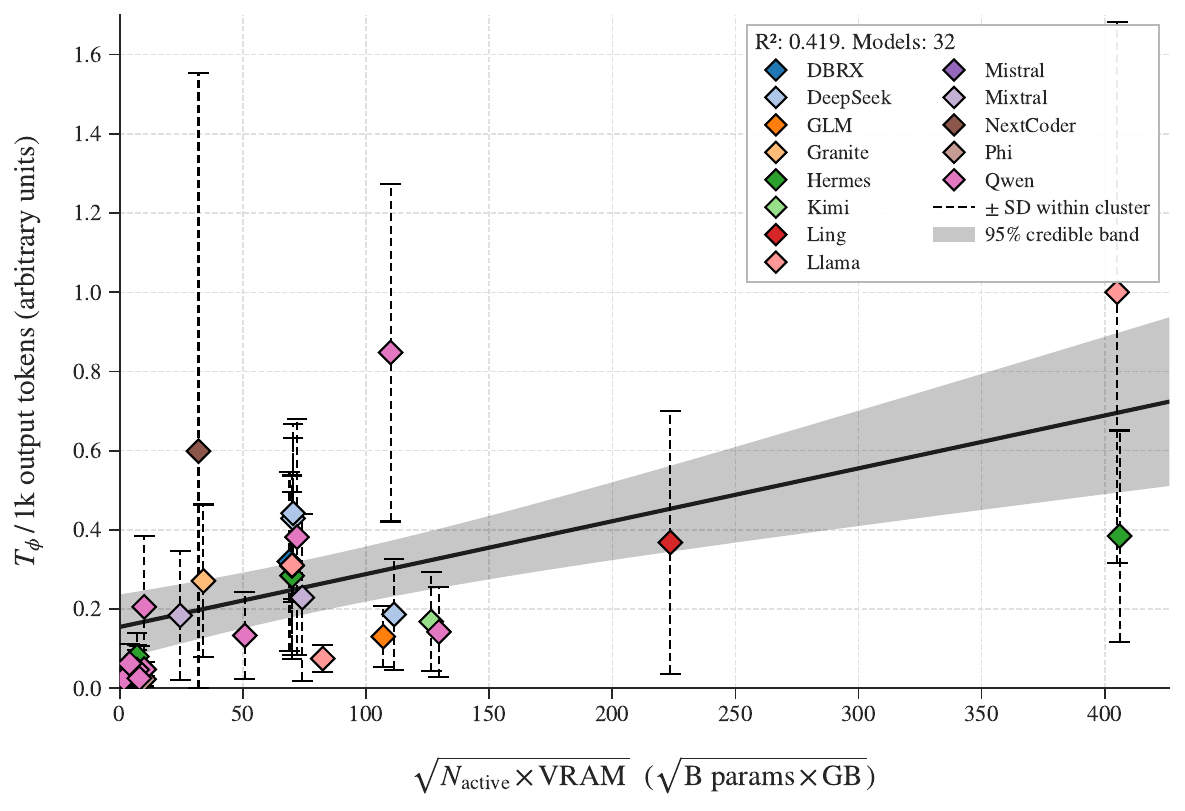}
  \caption{Decode-phase per-token wall clock time $T_{\phi}$ for all GPUs, quantisations and values of \npar, multiplied by a constant to mask the absolute scale.}
  \label{fig:walltime_decode}
\end{figure}

\begin{figure}[htbp]
  \centering
  \includegraphics[width=\figwidth]{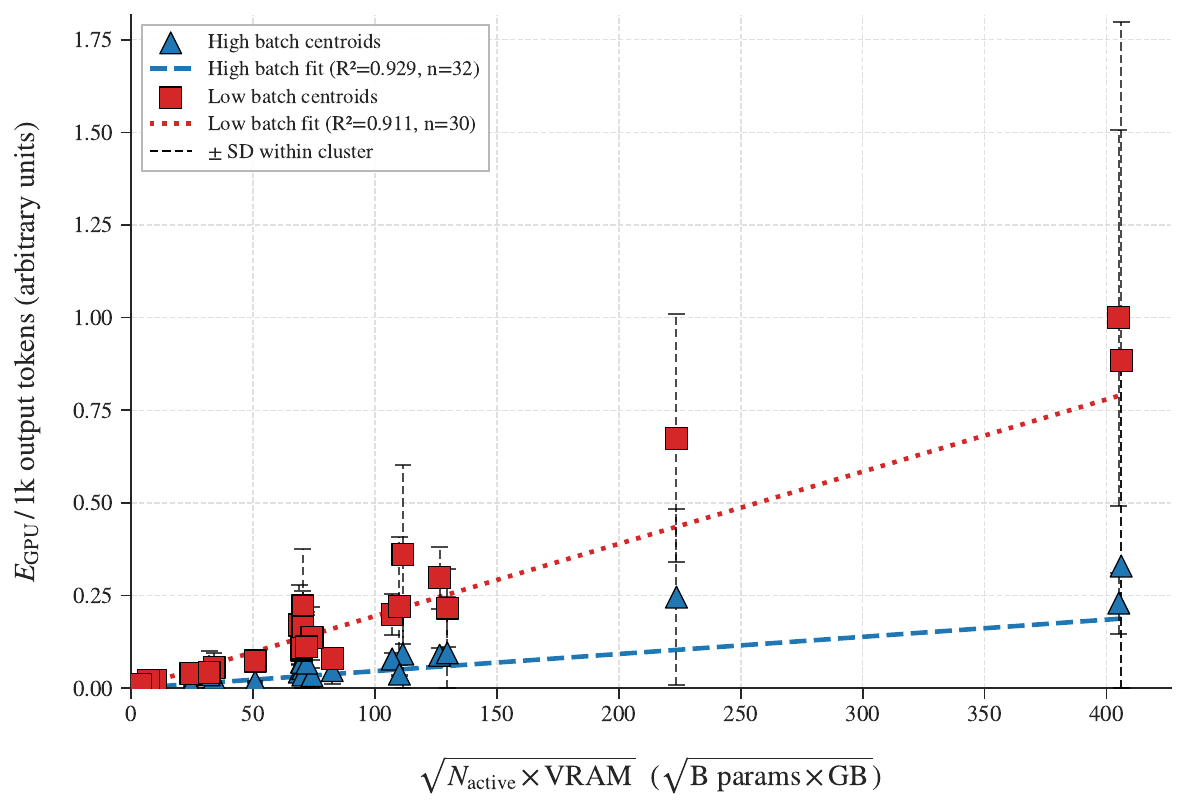}
  \caption{Effect of increasing \npar on the GPU energy regression for
  output tokens. Energy values have been multiplied by a constant to mask the absolute scale.}
  \label{fig:batching_overlay_output}
\end{figure}

Figures~\ref{fig:headline_kwh_input}-\ref{fig:walltime_decode} show the
regressions modelling GPU energy and wall clock time per input and output token
for the full benchmark dataset across all GPUs, quantisations and \npar values.
The 2D fitted regressions have been projected onto $\sqrt{N_{\mathrm{active}} \times \mathrm{VRAM}_{\mathrm{GB}}}$ for ease of illustration.
Notably, the per-token decode energy and wall clock time is found to be roughly an order of magnitude larger than prefill.
The exact multiplicative factor between the two token types cannot be inferred from the figures.
Coloured diamonds are per-model measurement centroids, one
per LLM and coloured by model family, with $\pm 1$ within-cluster standard
deviations (described in Section~\ref{sec:uncertainty_summary}) shown.
Figure~\ref{fig:batching_overlay_output} shows how decode energy per
token varies with the average number of parallel requests. We find no
consistent variation of energy per input token with batch size.

\begin{figure}[htbp]
  \centering
  \includegraphics[width=\figwidth]{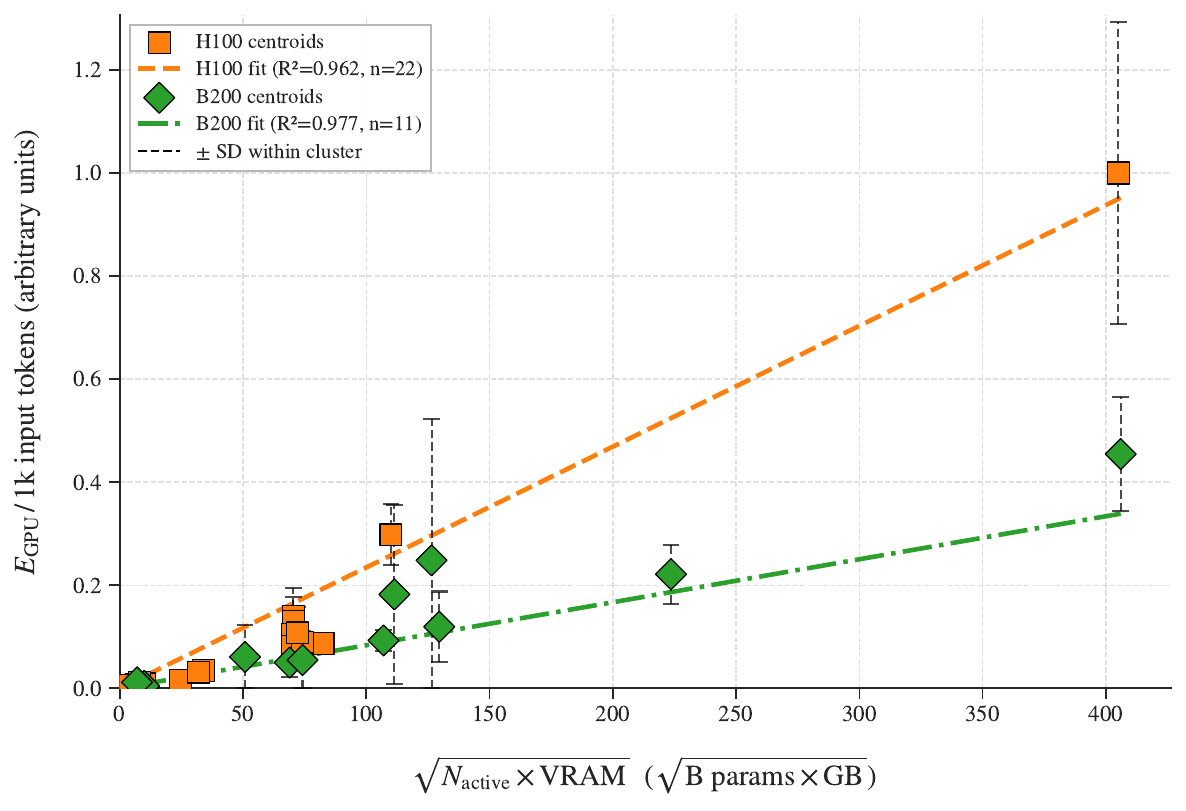}
  \caption{Effect of GPU model on the GPU energy
  regression for input tokens. Energy values have been multiplied by a constant to mask the absolute scale.}
  \label{fig:gpu_overlay_input}
\end{figure}

\begin{figure}[htbp]
  \centering
  \includegraphics[width=\figwidth]{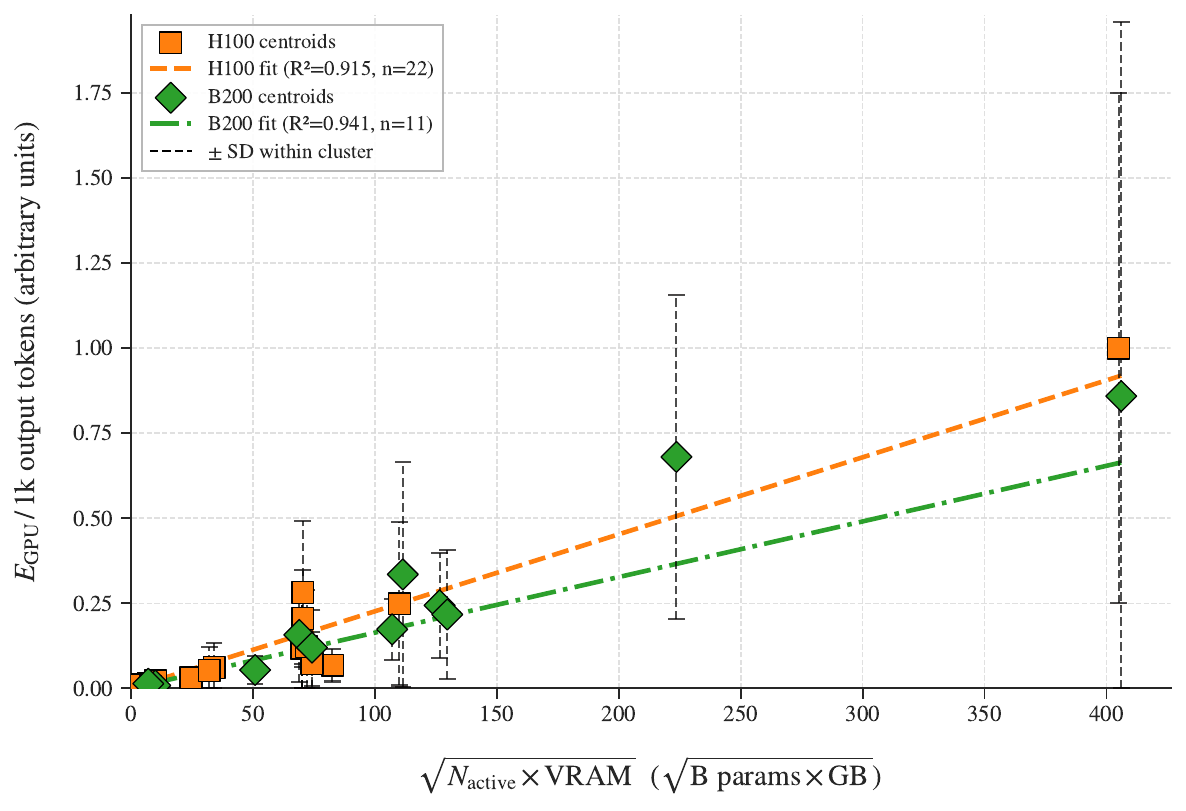}
  \caption{Effect of GPU model on the GPU energy
  regression for output tokens. Energy values have been multiplied by a constant to mask the absolute scale.}
  \label{fig:gpu_overlay_output}
\end{figure}

Figures~\ref{fig:gpu_overlay_input} and~\ref{fig:gpu_overlay_output}
split the regressions by GPU model. The newer B200 is more energy-efficient
than the H100 due to faster FLOP throughput, larger VRAM reserves and
memory bandwidth, and faster card-to-card link speeds, despite the
fact that the thermal design power (TDP) of the B200 far exceeds the H100.
Quantisations are pooled within each GPU series since NVFP4 versus
FP8 on the B200 did not significantly change the
position of the regression line.

\clearpage
\newpage

\begin{figure}[htbp]
  \centering
  \includegraphics[width=\figwidth]{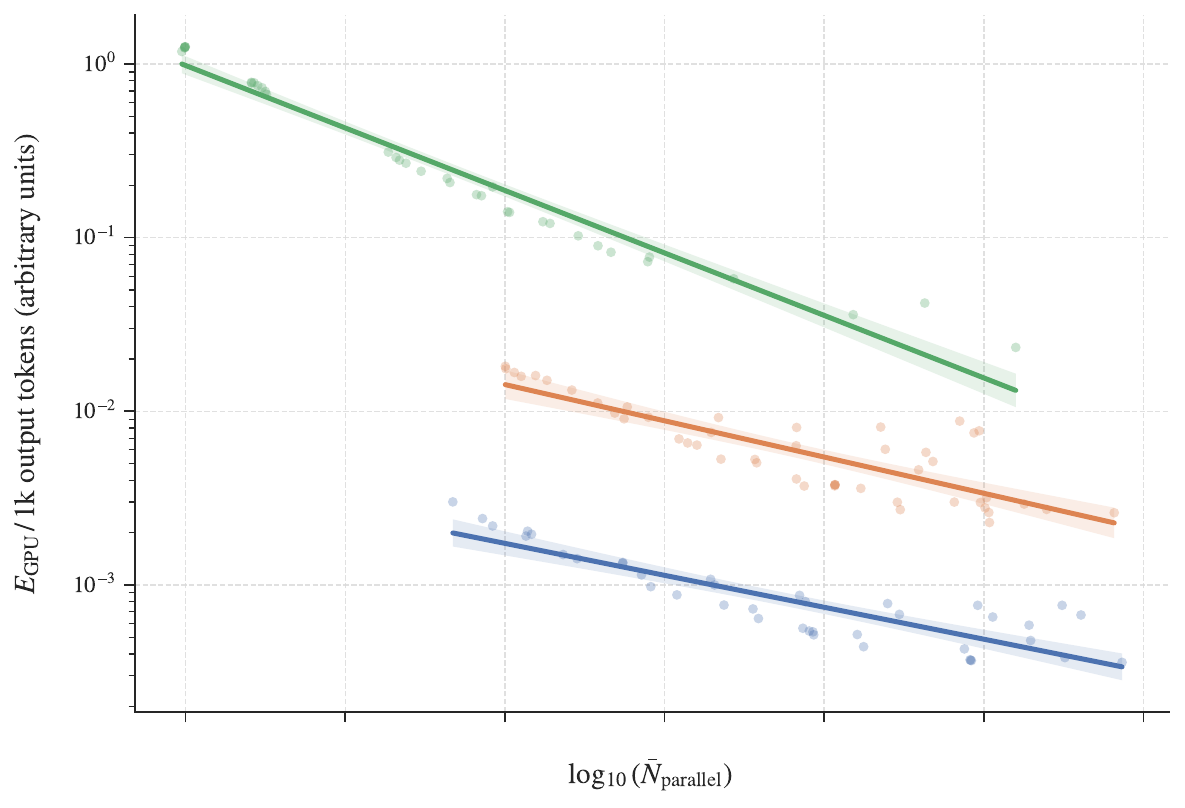}
  \caption{Relative GPU energy for $1000$ output tokens (multiplied by a constant to mask the absolute scale) versus
  $\log_{10}(\nparsym)$ (tick labels withheld) for three dense models, all on the H100 PCIe at FP8, each with
  a power-law fit.}
  \label{fig:parallel_output}
\end{figure}

\begin{figure}[htbp]
  \centering
  \includegraphics[width=\figwidth]{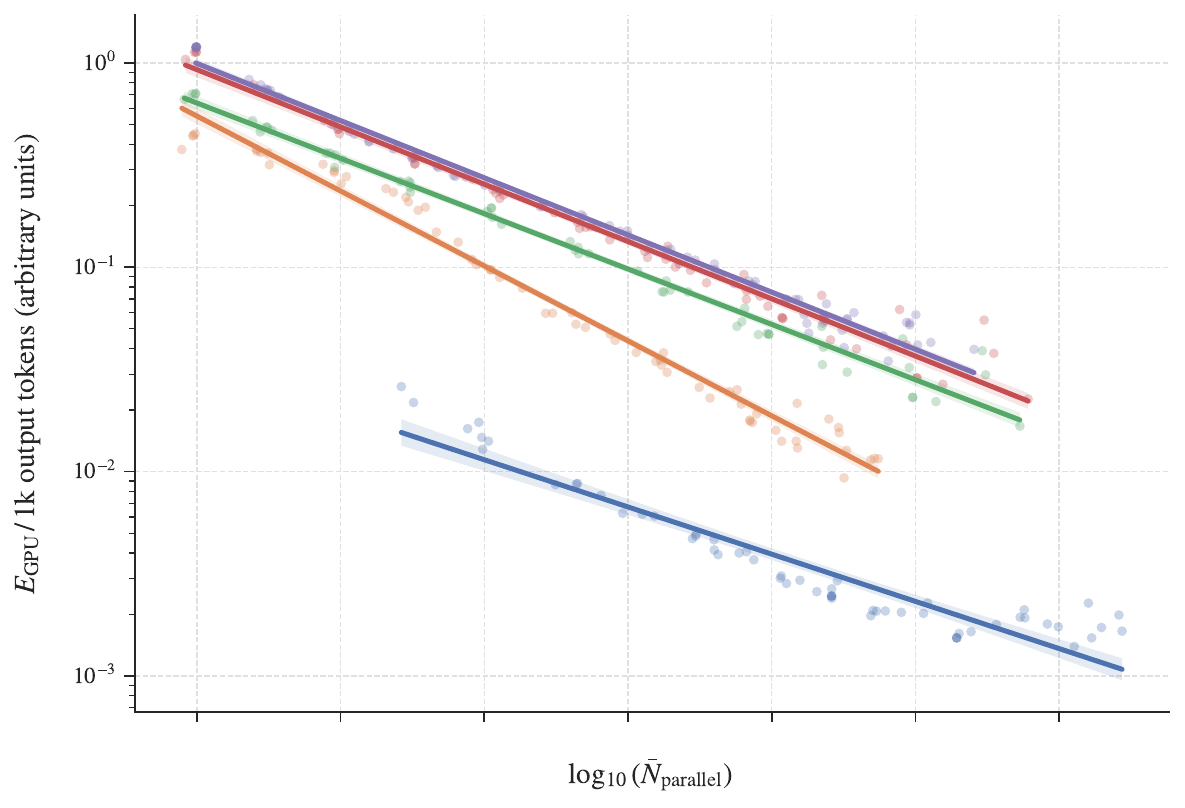}
  \caption{Relative GPU energy for $1000$ output tokens (multiplied by a constant to mask the absolute scale) versus
  $\log_{10}(\nparsym)$ (tick labels withheld) for five mixture-of-experts models, all on the B200.}
  \label{fig:parallel_moe_output}
\end{figure}

Figure~\ref{fig:parallel_output} shows, for three dense
models of different sizes, all
benchmarked on the same GPU (H100 PCIe) at FP8 quantisation, how per output token
energy scales with \npar. A straight line in log-linear space fits the data well, i.e., the decay is a power law
$E (\mathrm{kWh/token}) \propto \nparsym^{\,b}$.
This is approximately consistent with the exponential decay of energy per token with batch size
reported by Fernandez et al.~\citep{batchsize_energy_acl2025} and Samsi et
al.~\citep{samsi2023wordstowatts}. Figure~\ref{fig:parallel_moe_output} repeats
the exercise for five mixture-of-experts models, all benchmarked on the B200.
The same power-law decay holds.
Within the current dataset, the input (prefill) phase shows no consistent \npar scaling. It is flat
for the dense models, with at most a weak decay
in some MoE deployments on the B200.

These results show that a single-query figure quoted without
its serving context (e.g.\ Altman~\citep{altman2025gentlesingularity}) is
under-specified: the per-token energy cost can vary by more than an order of
magnitude across the realistic \npar range.

\clearpage
\newpage

\section{Mapping to proprietary models}
\label{sec:mapping}

\subsection{Naming-convention binning for proprietary tiers}
\label{sec:naming_hints}

Provider naming conventions are a soft indication of the number
of parameters in a previously-unseen proprietary model. We use these
conventions to assign each model to a parameter count
bin, which defines a minimum and maximum number of parameters the model
might have. These bins are named ``very large'', ``large'', ``medium'',
``small'' and ``very small''.
Each major provider uses a roughly consistent ordering of their model family.
We therefore assign binnings:

\begin{itemize}
  \item OpenAI: nano (very small); mini, including the o-series minis (small);
        o-series (large); GPT-5.x standard, Thinking and Pro (very large);
        GPT-5.6 Luna (small), Terra (medium) and Sol (large); GPT-6 Astra (very large).
  \item Anthropic: Haiku (small), Sonnet (medium), Opus (large), Fable (very large).
  \item Google: Flash-Lite (very small), Flash 3.x (small), Pro 3.x (large).
  \item xAI: Grok 4.x Fast (medium), Grok 4.x standard (large).
\end{itemize}

\subsection{Quantisation and VRAM assumption}
\label{sec:quantisation_assumption}

The deployed quantisation of proprietary models is not disclosed, but we assume
hyperscalers lean heavily on quantisation for several reasons:
\begin{itemize}
  \item FP8 has been the de facto
        production inference precision on Hopper-class GPUs since the
        introduction of the NVIDIA Transformer
        Engine~\citep{nvidia_hopper_fp8, micikevicius2022fp8}.
        However, Hopper architectures cannot support FP4 hosting \cite{nvidia_hopper_fp8}.
  \item When deploying on Blackwell architecture, NVFP4
        sacrifices marginal task-level performance for a substantial
        improvement in energy efficiency and a reduction in
        per-token latency~\citep{nvidia_blackwell_nvfp4}.
  \item Turboquant~\citep{turboquant} and other recent improvements in compression techniques have made aggressive low-bit
        deployment a competitive and cost-efficient option in hyperscalers.

\end{itemize}
The total VRAM footprint is the total parameter count $N_{\mathrm{total}}$ in
billions multiplied by the size of each parameter in bytes (0.5 for NVFP4 and 1 for FP8).
The active parameter count $N_{\mathrm{active}}$ scales with $N_{\mathrm{total}}$.

The number of cards per deployment is
chosen for each bin edge as the smallest power-of-two
that accommodates the model weights at vLLM's default $\texttt{gpu\_memory\_utilization}=0.9$,
with $\sim1.3\times$ uplift for KV cache, activations and
CUDA overhead. The required number of cards is therefore
\begin{equation}
      \label{eq:num_cards}
      N_{\mathrm{cards}} \simeq \mathrm{ceiling}\left(\frac{\mathrm{VRAM}_{\mathrm{GB}}}{\mathrm{VRAM}_{\mathrm{card}}\times(1/1.3)\times0.9}\right).
\end{equation}
These card counts in turn specify the non-GPU power via Table~\ref{tab:ec2_equivalent}
and scale the embodied emissions as detailed in Section~\ref{sec:emissions_water}.

\subsection{Expected value and uncertainty quantification via Monte Carlo sampling}
\label{sec:montecarlo}

For proprietary AI models we know at best a rough range for $N_{\mathrm{total}}$.
From this we assign a size bin and some feasible deployment configurations.
We do not make assumptions on the quantisation,
GPU card, \npar, training energy per token uplift, or the MoE fraction of
active parameters. Instead we assign a prior input distribution for each of these
unknowns and randomly sample them, reading off the regression prediction corresponding
to the sampled inputs. This produces a distribution of per-token energy and wall clock time
from which we obtain the median and interquartile range. This sampling enables us to
propagate uncertainty due to both continuous and discontinuous contributors, the latter including
non-GPU energy, which steps as a function of $N_{\mathrm{total}}$, card VRAM, and quantisation.
Taking the IQR as an uncertainty rather than a range per size bin reflects that
there are more degenerate ways to achieve energies closer to the median than the
tails of the distribution.

\begin{figure}[htbp]
  \centering
  \includegraphics[width=\textwidth]{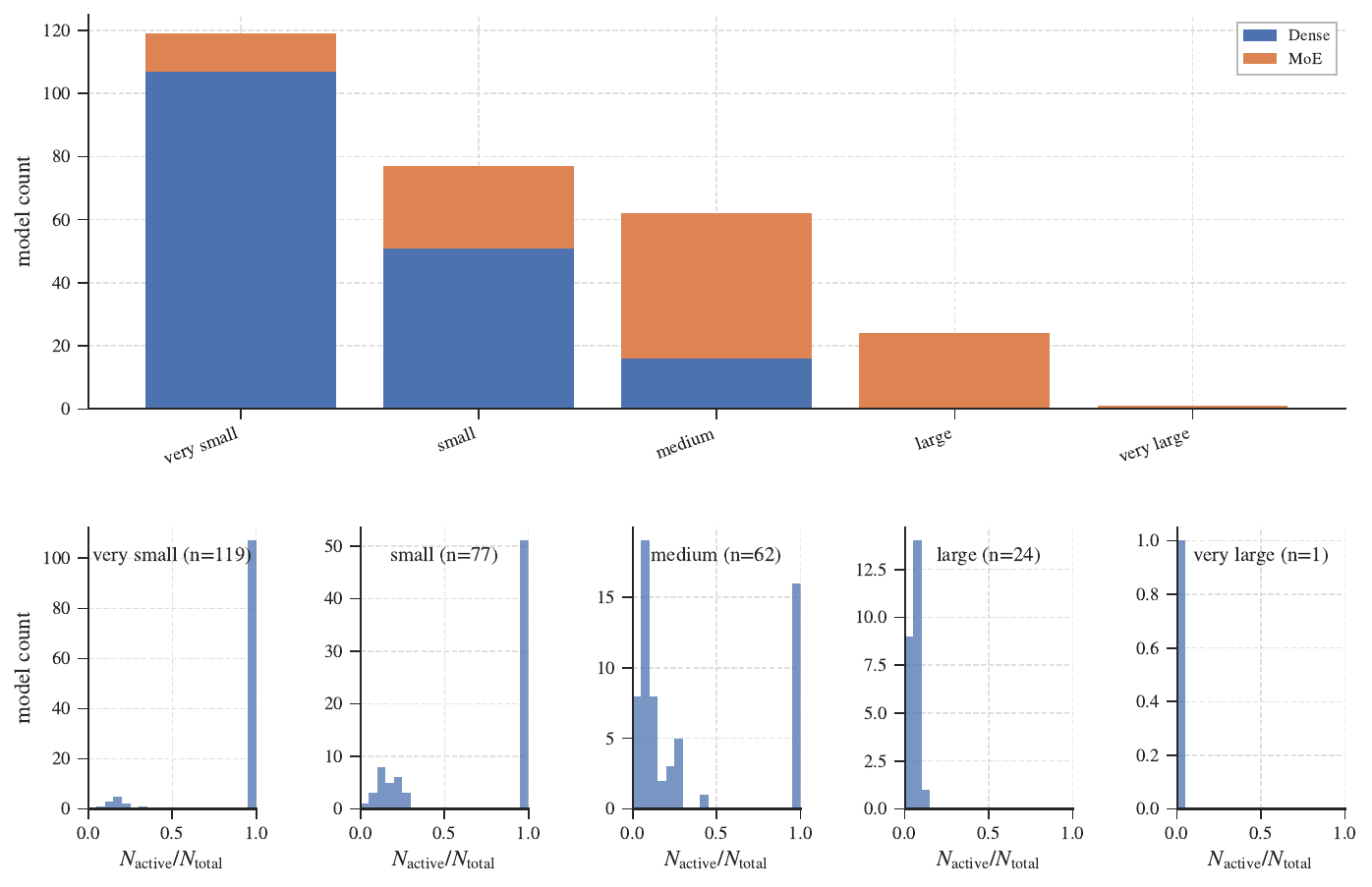}
  \caption{Top: dense versus
  mixture-of-experts counts per size bin. Bottom: the
  $N_{\mathrm{active}} / N_{\mathrm{total}}$ ratio distribution of the
  surveyed models in each size bin.}
  \label{fig:frontier_population}
\end{figure}

We estimate the prior probability that a model is dense ($N_{\mathrm{active}} = N_{\mathrm{total}}$)
as decreasing with model size bin: $80\,\%$ (very small), $60\,\%$ (small),
$20\,\%$ (medium) and $0\,\%$ (large and very large, where no dense models
exist to our knowledge), with $p_{\mathrm{mixture-of-experts}} = 1 - p_{\mathrm{dense}}$.
This is motivated by surveying $\sim250$ LLMs on HuggingFace from $25$
leading providers released from June 2023 onwards, with derivative
fine-tunes, distillations and quantised re-uploads excluded.
Figure~\ref{fig:frontier_population} shows the ratio of dense and MoE
architectures and the distribution of $N_{\mathrm{active}} / N_{\mathrm{total}}$
per size bin.

Since we have no prior knowledge on the likely training uplift, $N_{\mathrm{active}}$,
$N_{\mathrm{total}}$, hardware / quantisation and batching, we sample from
uniform prior distributions for these variables. In most cases, this is
a regular, linear uniform distribution, but $N_{\mathrm{active}}$
and $N_{\mathrm{total}}$ are log-uniform, such that every multiplicative
step in the range is equally likely. The hardware and
quantisation categories are: H100/FP8, B200/FP8, B200/NVFP4 (H100 does
not support NVFP4). The \npar categories are low, high
and a bin including all \npar values. The GPU card count is fixed by
the VRAM requirement of the sampled total parameters and quantisation.

The epistemic and aleatoric uncertainty of the regression fit is
propagated by sampling the fitted regression coefficients from the
posterior distribution and aleatoric variance. If a sampled configuration
corresponds to a decode wall-clock time per token that would give a
single-stream output throughput below $20$~tokens/s it is rejected as
implausible. API throughput rates tracked by Artificial
Analysis~\citep{artificial_analysis} and OpenRouter~\citep{openrouter}
rarely dip below this rate even for the largest frontier models.

\begin{figure}[p]
  \centering
  \includegraphics[width=\figwidth]{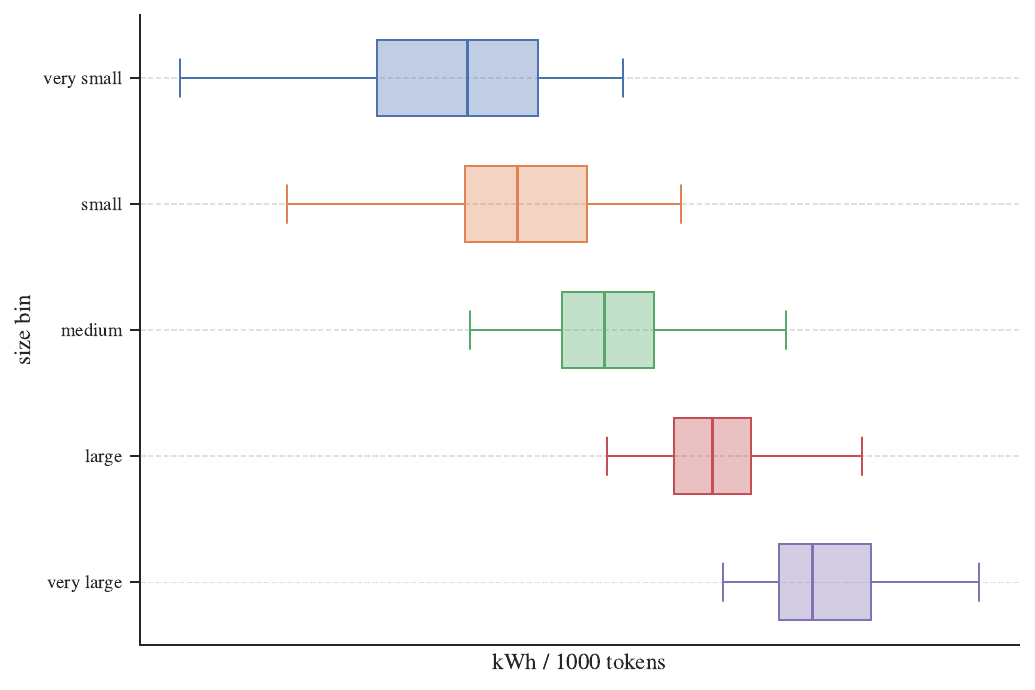}
  \caption{Sampled distribution of energy per input token.
  One box per $N_{\mathrm{total}}$ size bin in log energy per token
  space, with the energy axis ticks withheld to mask the absolute
  scale. Each box spans q25 to q75 with a line at the median; the
  whiskers reach the 1st and 99th percentiles.}
  \label{fig:energy_distribution_input}
\end{figure}

\begin{figure}[p]
  \centering
  \includegraphics[width=\figwidth]{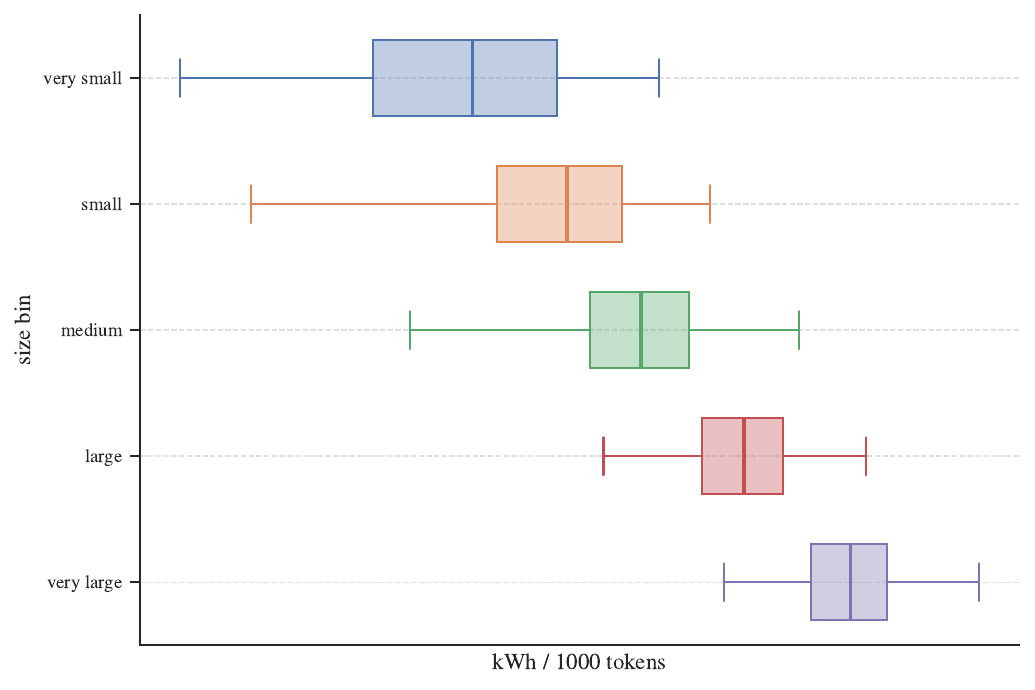}
  \caption{Sampled distribution of energy per output token.
  One box per $N_{\mathrm{total}}$ size bin in log energy per token
  space, with the energy axis ticks withheld to mask the absolute
  scale. Each box spans q25 to q75 with a line at the median; the
  whiskers reach the 1st and 99th percentiles.}
  \label{fig:energy_distribution_output}
\end{figure}

With the priors and uncertainty propagation defined, the sampling is
performed.
The final median and IQR server-level energy (including non-GPU components
without PUE applied) per input/output token,
conditioned on what a customer may actually know about
their deployment, form our estimate of the energy cost of closed-model inference in the public cloud.
Figures~\ref{fig:energy_distribution_input} and~\ref{fig:energy_distribution_output} illustrate the sampled distributions with no fixed priors.
As expected, the $N_{\mathrm{total}}$ size bins overlap marginally.
Wall clock time distributions formed in the same way feed into the embodied emissions predictions.
The median predictions for proprietary models are validated by
translating our energy per token values into an implied cost per token
at a representative regional electricity rate. If this cost is
greater than the API price per token being charged to the user,
the prediction is flagged for further review. Since the AI providers
may be currently running at a loss, this validation is not currently
definitive, but is nonetheless instructive.
Predicted per-token energy values have also been compared against
metered energy from enterprise Greenpixie clients running inference
on-premises. Discrepancies feed back into model refinement.

Sampling of the embodied \cotwoeq distribution is decoupled from the sampling
of the energy distribution. This is because, although
embodied emissions increase with energy per token within each $N_{\mathrm{total}}$ bin in our results,
that ordering is not guaranteed if the sampling is coupled. A dense model hosted on fewer cards can
use more energy per token than a mixture-of-experts model with a
small $N_{\mathrm{active}}$ spread over more cards, while carrying
lower embodied emissions per token.

\newpage

\subsection{Applicability to cloud and SaaS products}
\label{sec:applicability}

The per-token outputs are mapped onto cloud-service-provider SKUs by
model-name normalisation followed by lookup
against a reference table of this methodology's results. The join dimensions used
are

\begin{itemize}
  \item token type (input, output, embedding),
  \item batch tier (real-time, batched, priority, flexible),
  \item prefix-caching status triggered by a minimum hit-rate (cached vs.\ non-cached).
\end{itemize}

Each billing line is associated with the regression's per-token
GPU energy output and inference time, giving a median estimate and
an IQR uncertainty.
This means stakeholders can express the impact of their
choices with a credible spread rather than just a bare point estimate.
When regional information is available, regional hourly intensity
and PUE are applied separately per SKU.
Beyond cloud usage, the methodology is also directly
applicable to SaaS token-level usage reports, e.g.,\ OpenAI API, GitHub Copilot, Gemini, and enterprise
Anthropic per-token logs. Any report that exposes a
per-model token count by token type can be joined to the same
per-token kWh table.

\section{Discussion}
\label{sec:perf_emissions_tradeoff}

When considering what AI service to use for a given application,
an enterprise will make several FinOps and GreenOps architectural decisions.
Although not shown here directly, the results from the methodology show that
choosing the smallest LLM
capable of a task can improve the per-token energy efficiency by
orders of magnitude, making the model choice within a family/provider
one of the largest emissions levers available. Enterprises should
therefore consider routing low-stakes requests (classification,
summarisation, formatting) to a small-tier LLM and reserve
the larger tiers for high-stakes reasoning. Additionally, it is
important to consider ``tokens per task'': i.e., for two similarly
sized LLMs, how many extra tokens and therefore extra joules of
energy does one LLM use over another? This is benchmarked on
Artificial Analysis \citep{artificial_analysis}.

The results here also give strong motivation to consider batched
SKU line items offered by the cloud providers. At the cost of some
latency in building a batch, processing more requests in parallel utilises
the GPU much more efficiently, leading to far less operational energy
per token for the same model. Workloads that can tolerate longer
turnaround (offline summarisation, document classification) should run
batched.

These findings go hand-in-hand with other best practices in
sustainable AI usage. Shorter system prompts, more aggressive
output-length limits, and avoiding chain-of-thought traces where
they do not measurably improve accuracy all help to reduce the total
number of tokens. Clever prompt engineering that can trade
output tokens for input tokens will save roughly an order of magnitude
in energy per token. Caching of input tokens to avoid repeated prefill
computations~\citep{wang2025storellm} saves both cost and energy,
and is available for most major providers. Larger architecture decisions,
such as the region the AI workloads run in, can also make an enormous
difference. For example, across the AWS / Azure / GCP regions that
Greenpixie supports, \cotwoeq-intensity per kWh
varies by more than $30\times$.

Enterprise decision makers are likely to consider model quality, latency,
and unit cost when making decisions on their AI cloud infrastructure.
These metrics form a Pareto frontier with Greenpixie's per-token
sustainability metrics which serve to inform stakeholder GreenOps decisions.
Cost often correlates with the environmental impact of AI,
so improving one is likely to improve the other.

\subsection{Limitations and future work}
\label{sec:limitations}

We believe that this methodology provides robust and actionable numbers that can be
used to help enterprises make sustainable choices in the majority of LLM use cases.
The methodology is being progressed towards International Organization
for Standardization (ISO) compliance, and its scope is widened on a continuous basis.
AI is a fast-moving field, so future maintenance work is planned to
ensure that the benchmarking configuration, deployment hardware, open-weights LLMs,
benchmarking tasks and proprietary model sizing all remain relevant and up-to-date
with the frontier. At the time of writing, the open-weights frontier has already
moved since this campaign closed - benchmarking the latest models is a high short-term priority.
In particular, DeepSeek-V4-Pro/V4.1-Flash and Kimi K3 will extend our model coverage past 1T parameters,
improving our extrapolation to the largest proprietary models. Any estimate for LLMs
occupying more than 1.5~TB of VRAM is an extrapolation from our current dataset.
Future work will involve building an automated benchmarking pipeline to
test new frontier open-weights models within days of their release, and to triage
new proprietary models and assign them to parameter count bins automatically.

An unavoidable limitation is the uncertainty related to the unknown deployment
configuration and architecture of proprietary LLMs. Assignment of a proprietary
model to the incorrect parameter-count bin would make a large difference to the
estimates. A related limitation is that some cloud and SaaS products
advertise one LLM but internally route requests to a family of LLMs depending on
request content (e.g.\ a router that escalates difficult
requests to a larger model, or fall-back routing under
capacity pressure). Our predictions for such products represent the advertised LLM only.

Future work will include exploration of different features that correlate
with energy per token and per task. For open-weights models with publicly-available
architectural detail, a richer regression model with more features
should give materially tighter predictions for cloud deployments than the
two-feature regression model used here. Provider-reported token throughput
could potentially be explored as a proxy for deployment hardware and
model architecture in the style of EcoLogits~\citep{ecologits} and
Jegham et al.~\citep{jegham2025hungryaibenchmarkingenergy}.

Our benchmarking setup has a number of limitations. GPU rental power draw
on Runpod for $\leq 8$-GPU configurations is assumed to be representative
of equivalent hyperscaler racks. Runpod does not publish chassis, PUE,
or cooling specifications, so this holds at the GPU-board level only.
Rack-level and cluster-level effects are not captured.
We note specifically that the NVIDIA GB200 NVL72 rack
delivers $1.8$~TB/s per-GPU and an aggregate $130$~TB/s
NVLink Switch bandwidth~\citep{nvidia_dgx_gb200_nvl72},
roughly an order of magnitude greater than a PCIe~Gen5 interconnect.
For 1T+ parameter LLMs that would traditionally span multiple nodes,
the GB200 NVL72 avoids inter-node hops that the Runpod 8-GPU baseline incurs, so
very-large LLMs may be much more energy efficient than this methodology would suggest.
We also assume that the number of identical cards required to host an LLM is the minimum power of
two that will satisfy the LLM's VRAM requirements. This may not always be the case, since some
providers may want a larger memory and parallelism capacity, or could
run using pipeline parallelism instead of tensor parallelism.
Benchmarking across a wider range of GPU cards (and non-GPU components)
from different vendors in different configurations would also reduce the
uncertainty in cases where the deployment is known, and likely increase the uncertainties
assigned here on unknown deployments. Further research and development is
planned to improve our approximation of continuous batching and extend
our LLM request-serving to multiple parallel multi-turn conversations.

There are also a variety of ways that our work could be expanded upon to cover
different AI use cases. Image, audio and video generation should be included.
Static-batched flexible deployments, single-request real-time deployments,
and provider-specific configurations (speculative decoding etc.) are not
directly represented. Different tokenizers lead to a different number
of tokens for a given sentence, affecting the energy per token subtly.

A key limitation that is in scope to address very soon is the
extension to modern agentic workflows. Agentic and long-context
workloads can run at up to $1{,}000{,}000$ tokens per prompt, which
may have a significant effect on the energy efficiency per input and
output token. Longer prompts may also reveal a trend in input token energy
as a function of \npar. Longer multi-turn sessions (e.g. in software
engineering) make heavy usage of prefix caching~\citep{zhu2026tracelab}, which
is enabled in our configuration but its impact on energy per token
is not yet thoroughly explored. The impact of agentic tool usage
(web search, code execution, retrieval-augmented generation,
structured-output validators) should ideally be quantified and
amortised back onto a per-token basis so that agentic-workflow
estimates can be reported alongside the raw LLM-inference numbers.

\section{Conclusion}
\label{sec:conclusion_summary}
This methodology provides per-token, server-level energy estimates
for cloud AI inference separately for prefill and decode tokens
and propagates them through to \cotwoeq and water estimates using the
operational and embodied factors from existing Greenpixie methodologies.
The regression model is trained on a benchmark dataset of open-weights LLMs from
1.8B to 1T total parameters on Hopper and Blackwell-class hardware
at production-realistic quantisations and batch sizes.

The prediction for an unbenchmarked proprietary model is estimated using MC sampling of feasible
deployment configurations within a range of model parameter counts.
These inputs to the regression model are sampled from prior distributions, and the regression's own
posterior uncertainty is propagated into every sample.
This quantifies the uncertainty of the unknown parameter count, hardware, quantisation, \npar and training energy cost.

The output is designed to be actionable, and honest in terms of uncertainty. Greenpixie uses this methodology
of per-SKU kWh, g\cotwoeq and litres of water
per 1000 tokens to enrich every cloud line item associated with LLM inference that an
enterprise customer is billed for. In this way, business decisions can be made on sustainability grounds
alongside cost, quality, and latency.

\section*{Generative AI usage statement}
The authors used Claude (Anthropic; various versions) to assist with writing
the analysis and plotting code, exploratory data analysis, and
editing the manuscript. All outputs were reviewed, tested and edited by the authors,
who take full responsibility for the content of this paper.

\clearpage
\bibliographystyle{unsrtnat}
\bibliography{whitepaper}

@inproceedings{vllm_kwon2023,
  title     = {Efficient Memory Management for Large Language Model Serving with {PagedAttention}},
  author    = {Kwon, Woosuk and others},
  booktitle = {Proceedings of the 29th Symposium on Operating Systems Principles (SOSP)},
  year      = {2023},
}

@inproceedings{wang2025storellm,
  title     = {{StoreLLM}: Energy Efficient Large Language Model Inference with Permanently Pre-stored Attention Matrices},
  author    = {Wang, Dan and others},
  booktitle = {Proceedings of the 16th ACM International Conference on Future and Sustainable Energy Systems (E-Energy)},
  year      = {2025},
  doi       = {10.1145/3679240.3734604},
}

@misc{micikevicius2022fp8,
  title  = {{FP8} Formats for Deep Learning},
  author = {Micikevicius, Paulius and others},
  year   = {2022},
  howpublished = {arXiv:2209.05433},
}

@misc{nvidia_hopper_fp8,
  title  = {{NVIDIA} {Hopper} Architecture In-Depth},
  author = {{NVIDIA}},
  year   = {2022},
  howpublished = {\url{https://developer.nvidia.com/blog/nvidia-hopper-architecture-in-depth/}},
}

@misc{aws_accelerated_computing,
  title  = {Amazon {EC2} Accelerated Computing Instances},
  author = {{Amazon Web Services}},
  year   = {2026},
  howpublished = {\url{https://aws.amazon.com/ec2/instance-types/accelerated-computing/}},
}

@misc{turboquant,
  title  = {{TurboQuant}: Online Vector Quantization with Near-optimal Distortion Rate},
  author = {Zandieh, Amir and others},
  year   = {2025},
  howpublished = {arXiv:2504.19874},
}

@article{henderson2020systematic,
  title   = {Towards the Systematic Reporting of the Energy and Carbon Footprints of Machine Learning},
  author  = {Henderson, Peter and others},
  journal = {Journal of Machine Learning Research},
  volume  = {21},
  number  = {248},
  pages   = {1--43},
  year    = {2020},
}

@inproceedings{wu2022sustainable,
  title     = {Sustainable {AI}: Environmental Implications, Challenges and Opportunities},
  author    = {Wu, Carole-Jean and others},
  booktitle = {Proceedings of Machine Learning and Systems},
  volume    = {4},
  year      = {2022},
}

@article{desislavov2023inference,
  title   = {Trends in {AI} inference energy consumption: Beyond the performance-vs-parameter laws of deep learning},
  author  = {Desislavov, Radosvet and Mart{\'i}nez-Plumed, Fernando and Hern{\'a}ndez-Orallo, Jos{\'e}},
  journal = {Sustainable Computing: Informatics and Systems},
  volume  = {38},
  pages   = {100857},
  year    = {2023},
}

@misc{trott2024tokenization,
  title  = {Tokenization in large language models, explained},
  author = {Trott, Sean},
  year   = {2024},
  howpublished = {The Counterfactual. \url{https://seantrott.substack.com/p/tokenization-in-large-language-models}},
}

@misc{llm_inference_handbook_prefill_decode,
  title  = {Prefill-decode disaggregation},
  author = {{Modular}},
  year   = {2026},
  howpublished = {LLM Inference Handbook. \url{https://handbook.modular.com/inference-optimization/prefill-decode-disaggregation/}},
}

@misc{llm_inference_handbook_batching,
  title  = {Static, dynamic and continuous batching},
  author = {{Modular}},
  year   = {2026},
  howpublished = {LLM Inference Handbook. \url{https://handbook.modular.com/inference-optimization/static-dynamic-continuous-batching}},
}

@misc{goldman_token_growth_2026,
  title  = {{AI} Agents Forecast to Boost Tech Cash Flow as Usage Soars},
  author = {{Goldman Sachs}},
  year   = {2026},
  howpublished = {\url{https://www.goldmansachs.com/insights/articles/ai-agents-forecast-to-boost-tech-cash-flow-as-usage-soars}},
}

@misc{linuxfoundation_tokenomics_2026,
  title  = {{Linux Foundation} Launches the {Tokenomics Foundation} to Define the Economics and {ROI} of {AI} Value},
  author = {{Linux Foundation}},
  year   = {2026},
  howpublished = {\url{https://www.linuxfoundation.org/press/linux-foundation-launches-the-tokenomics-foundation-to-define-the-economics-and-roi-of-ai-value}},
}

@misc{un_ai_environment_2026,
  title  = {{AI}'s environmental costs threaten water, land and climate},
  author = {{UN News}},
  year   = {2026},
  howpublished = {\url{https://news.un.org/en/story/2026/06/1167658}},
}

@misc{runpod_docs,
  title  = {{API} Reference},
  author = {{Runpod}},
  year   = {2026},
  howpublished = {\url{https://docs.runpod.io/api-reference/overview}},
}

@misc{jegham2025hungryaibenchmarkingenergy,
  title  = {How Hungry is {AI}? {B}enchmarking Energy, Water, and Carbon Footprint of {LLM} Inference},
  author = {Jegham, Nidhal and others},
  year   = {2025},
  howpublished = {arXiv:2505.09598},
}

@misc{epoch_chatgpt_energy_2025,
  title  = {How much energy does {ChatGPT} use?},
  author = {You, Josh},
  year   = {2025},
  howpublished = {Epoch AI Gradient Updates. \url{https://epoch.ai/gradient-updates/how-much-energy-does-chatgpt-use}},
}

@misc{samsi2023wordstowatts,
  title  = {From Words to Watts: Benchmarking the Energy Costs of Large Language Model Inference},
  author = {Samsi, Siddharth and others},
  year   = {2023},
  howpublished = {arXiv:2310.03003},
}

@misc{altman2025gentlesingularity,
  title  = {The Gentle Singularity},
  author = {Altman, Sam},
  year   = {2025},
  howpublished = {\url{https://blog.samaltman.com/the-gentle-singularity}},
}

@misc{stojkovic2024llmenergy,
  title  = {Towards Greener {LLM}s: Bringing Energy-Efficiency to the Forefront of {LLM} Inference},
  author = {Stojkovic, Jovan and others},
  year   = {2024},
  howpublished = {arXiv:2403.20306},
}

@inproceedings{batchsize_energy_acl2025,
  title     = {Energy Considerations of Large Language Model Inference and Efficiency Optimizations},
  author    = {Fernandez, Jared and others},
  booktitle = {Proceedings of the 63rd Annual Meeting of the Association for Computational Linguistics (ACL)},
  year      = {2025},
}

@misc{openrouter,
  title  = {{LLM} Router and Marketplace},
  author = {{OpenRouter}},
  year   = {2026},
  howpublished = {\url{https://openrouter.ai}},
}

@misc{ecologits,
  title  = {{EcoLogits}: Estimating the Environmental Impact of generative {AI} models},
  author = {{GenAI Impact}},
  year   = {2024},
  howpublished = {\url{https://ecologits.ai}},
}

@inproceedings{chung2025mlenergy_benchmark,
  title     = {The {ML.ENERGY} Benchmark: Toward Automated Inference Energy Measurement and Optimization},
  author    = {Chung, Jae-Won and others},
  booktitle = {Advances in Neural Information Processing Systems (NeurIPS), Datasets and Benchmarks Track},
  year      = {2025},
  note      = {arXiv:2505.06371},
}

@misc{chung2026joules,
  title  = {Where Do the Joules Go? {D}iagnosing Inference Energy Consumption},
  author = {Chung, Jae-Won and others},
  year   = {2026},
  howpublished = {arXiv:2601.22076},
}

@inproceedings{wilkins2024offline_energy,
  title     = {Offline Energy-Optimal {LLM} Serving: Workload-Based Energy Models for {LLM} Inference on Heterogeneous Systems},
  author    = {Wilkins, Grant and others},
  booktitle = {Workshop on Sustainable Computer Systems (HotCarbon)},
  year      = {2024},
  note      = {arXiv:2407.04014},
}

@inproceedings{niu2026tokenpowerbench,
  title     = {{TokenPowerBench}: Benchmarking the Power Consumption of {LLM} Inference},
  author    = {Niu, Chenxu and others},
  booktitle = {Proceedings of the AAAI Conference on Artificial Intelligence},
  year      = {2026},
  note      = {arXiv:2512.03024},
}

@misc{vartziotis2026tokens_to_wh,
  title  = {From Tokens to Watt-hours: Analytical Energy Estimation for {LLM} Inference on Modern {GPU}s},
  author = {Vartziotis, Tina and others},
  year   = {2026},
  howpublished = {arXiv:2607.26571},
}

@inproceedings{vellaisamy2026token_energy,
  title     = {Characterization of Request and Token Energy Costs for {LLM} Inference Workloads on {GPU} Platforms},
  author    = {Vellaisamy, Prabhu and others},
  booktitle = {IEEE International Symposium on Workload Characterization (IISWC)},
  year      = {2026},
  note      = {arXiv:2608.28044},
}

@inproceedings{lottick2019codecarbon,
  title  = {Energy Usage Reports: Environmental awareness as part of algorithmic accountability},
  author = {Lottick, Kadan and others},
  booktitle = {Workshop on Tackling Climate Change with Machine Learning, NeurIPS},
  year   = {2019},
}

@misc{luccioni2024power_hungry,
  title  = {Power Hungry Processing: Watts Driving the Cost of {AI} Deployment?},
  author = {Luccioni, Alexandra Sasha and Jernite, Yacine and Strubell, Emma},
  year   = {2024},
  howpublished = {arXiv:2311.16863},
}

@misc{ai_energy_score_2024,
  title  = {{AI} {E}nergy {S}core: A Leaderboard for measuring {AI} model energy efficiency},
  author = {{Hugging Face and Salesforce}},
  year   = {2024},
  howpublished = {\url{https://huggingface.co/AIEnergyScore}},
}

@article{luccioni2023bloom,
  title   = {Estimating the Carbon Footprint of {BLOOM}, a 176{B} Parameter Language Model},
  author  = {Luccioni, Alexandra Sasha and Viguier, Sylvain and Ligozat, Anne-Laure},
  journal = {Journal of Machine Learning Research},
  volume  = {24},
  number  = {253},
  pages   = {1--15},
  year    = {2023},
}

@misc{mistral_large_2_lca,
  title  = {Our Contribution to a Global Environmental Standard for {AI}},
  author = {{Mistral AI}},
  year   = {2025},
  howpublished = {\url{https://mistral.ai/news/our-contribution-to-a-global-environmental-standard-for-ai/}},
}

@misc{google_gemini_emissions_2025,
  title  = {Measuring the Environmental Impact of Delivering {AI} at {Google} Scale},
  author = {{Google}},
  year   = {2025},
  howpublished = {\url{https://services.google.com/fh/files/misc/measuring_the_environmental_impact_of_delivering_ai_at_google_scale.pdf}},
}

@misc{nvidia_h100_pcie,
  title  = {{NVIDIA} {H100} {Tensor Core} {GPU}},
  author = {{NVIDIA}},
  year   = {2023},
  howpublished = {\url{https://www.nvidia.com/en-us/data-center/h100/}},
}

@misc{nvidia_b200,
  title  = {{NVIDIA} {Blackwell} {B200} {Tensor Core} {GPU}},
  author = {{NVIDIA}},
  year   = {2024},
  howpublished = {\url{https://www.nvidia.com/en-gb/data-center/technologies/blackwell-architecture/}},
}

@misc{nvidia_dgx_gb200_nvl72,
  title  = {{NVIDIA} {GB200} {NVL72}},
  author = {{NVIDIA}},
  year   = {2024},
  howpublished = {\url{https://www.nvidia.com/en-us/data-center/gb200-nvl72/}},
}

@misc{nvidia_blackwell_nvfp4,
  title  = {Introducing {NVFP4} for Efficient and Accurate Low-Precision Inference},
  author = {{NVIDIA}},
  year   = {2025},
  howpublished = {\url{https://developer.nvidia.com/blog/introducing-nvfp4-for-efficient-and-accurate-low-precision-inference/}},
}

@misc{artificial_analysis,
  title  = {Independent Analysis of {AI} Models and Providers},
  author = {{Artificial Analysis}},
  year   = {2026},
  howpublished = {\url{https://artificialanalysis.ai/}},
}

@misc{nvidia_nvml,
  title  = {{NVIDIA} Management Library ({NVML}) Reference},
  author = {{NVIDIA}},
  year   = {2024},
  howpublished = {\url{https://docs.nvidia.com/deploy/nvml-api/}},
}

@misc{zhou2023ifeval,
  title  = {Instruction-Following Evaluation for Large Language Models},
  author = {Zhou, Jeffrey and others},
  year   = {2023},
  howpublished = {arXiv:2311.07911},
}

@misc{taori2023alpaca,
  title  = {Stanford {A}lpaca: An Instruction-following {LL}a{MA} model},
  author = {Taori, Rohan and others},
  year   = {2023},
  howpublished = {\url{https://github.com/tatsu-lab/stanford_alpaca}},
}

@misc{suzgun2022bbh,
  title  = {Challenging {BIG}-{B}ench Tasks and Whether Chain-of-Thought Can Solve Them},
  author = {Suzgun, Mirac and others},
  year   = {2022},
  howpublished = {arXiv:2210.09261},
}

@misc{cobbe2021gsm8k,
  title  = {Training Verifiers to Solve Math Word Problems},
  author = {Cobbe, Karl and others},
  year   = {2021},
  howpublished = {arXiv:2110.14168},
}

@misc{clark2018arc,
  title  = {Think you have Solved Question Answering? {T}ry {ARC}, the {AI2} Reasoning Challenge},
  author = {Clark, Peter and others},
  year   = {2018},
  howpublished = {arXiv:1803.05457},
}

@misc{phan2025hle,
  title  = {Humanity's Last Exam},
  author = {Phan, Long and others},
  year   = {2025},
  howpublished = {arXiv:2501.14249},
}

@misc{rein2023gpqa,
  title  = {{GPQA}: A Graduate-Level {Google}-Proof {Q}\&{A} Benchmark},
  author = {Rein, David and others},
  year   = {2023},
  howpublished = {arXiv:2311.12022},
}

@misc{wang2024mmlupro,
  title  = {{MMLU-Pro}: A More Robust and Challenging Multi-Task Language Understanding Benchmark},
  author = {Wang, Yubo and others},
  year   = {2024},
  howpublished = {arXiv:2406.01574},
}

@misc{chen2021humaneval,
  title  = {Evaluating Large Language Models Trained on Code},
  author = {Chen, Mark and others},
  year   = {2021},
  howpublished = {arXiv:2107.03374},
}

@inproceedings{rajpurkar2018squadv2,
  title     = {Know What You Don't Know: Unanswerable Questions for {SQuAD}},
  author    = {Rajpurkar, Pranav and Jia, Robin and Liang, Percy},
  booktitle = {Proceedings of the 56th Annual Meeting of the Association for Computational Linguistics (ACL)},
  year      = {2018},
}

@inproceedings{hermann2015cnndaily,
  title     = {Teaching Machines to Read and Comprehend},
  author    = {Hermann, Karl Moritz and others},
  booktitle = {Advances in Neural Information Processing Systems (NeurIPS)},
  year      = {2015},
}

@inproceedings{narayan2018xsum,
  title     = {Don't Give Me the Details, Just the Summary! {T}opic-Aware Convolutional Neural Networks for Extreme Summarization},
  author    = {Narayan, Shashi and Cohen, Shay B. and Lapata, Mirella},
  booktitle = {Proceedings of the 2018 Conference on Empirical Methods in Natural Language Processing (EMNLP)},
  year      = {2018},
}

@inproceedings{maas2011imdb,
  title     = {Learning Word Vectors for Sentiment Analysis},
  author    = {Maas, Andrew L. and others},
  booktitle = {Proceedings of the 49th Annual Meeting of the Association for Computational Linguistics},
  year      = {2011},
}

@misc{merity2016wikitext,
  title  = {Pointer Sentinel Mixture Models},
  author = {Merity, Stephen and others},
  year   = {2016},
  howpublished = {arXiv:1609.07843},
}

@misc{patterson2023sciencedirect,
  title  = {The Carbon Footprint of Machine Learning Training Will Plateau, Then Shrink},
  author = {Patterson, David and others},
  year   = {2022},
  howpublished = {arXiv:2204.05149},
}

@misc{aws_inferentia_2019,
  title  = {Deliver high-performance {ML} inference with {AWS} {Inferentia}},
  author = {{Amazon Web Services}},
  year   = {2019},
  howpublished = {\url{https://d1.awsstatic.com/events/reinvent/2019/REPEAT_1_Deliver_high_performance_ML_inference_with_AWS_Inferentia_CMP324-R1.pdf}},
}

@misc{nvidia_hpcwire_2019,
  title  = {{AWS} Upgrades Its {GPU}-Backed {AI} Inference Platform},
  author = {{HPCwire}},
  year   = {2019},
  howpublished = {\url{https://www.hpcwire.com/2019/03/19/aws-upgrades-its-gpu-backed-ai-inference-platform/}},
}

@misc{mit_tech_review_2025,
  title  = {We Did the Math on {AI}'s Energy Footprint. {H}ere's the Story You Haven't Heard.},
  author = {{MIT Technology Review}},
  year   = {2025},
  howpublished = {\url{https://www.technologyreview.com/2025/05/20/1116327/ai-energy-usage-climate-footprint-big-tech/}},
}

@misc{boavizta_server,
  title  = {Methodologies for Digital Environmental Impact Assessment},
  author = {{Boavizta}},
  year   = {2024},
  howpublished = {\url{https://boavizta.org/en/methodologies}},
}

@misc{boavizta_api_manufacture,
  title  = {Boavizta {API}: Embedded Impacts Methodology},
  author = {{Boavizta}},
  year   = {2024},
  howpublished = {\url{https://doc.api.boavizta.org/Explanations/embedded_methodology/}},
}

@article{li2023making_ai_less_thirsty,
  title   = {Making {AI} Less ``Thirsty'': Uncovering and Addressing the Secret Water Footprint of {AI} Models},
  author  = {Li, Pengfei and others},
  journal = {Communications of the ACM},
  volume  = {68},
  number  = {7},
  pages   = {54--61},
  year    = {2025},
  doi     = {10.1145/3724499},
}

@misc{wri_water_electricity_2020,
  title  = {Guidance for Calculating Water Use Embedded in Purchased Electricity},
  author = {Reig, Paul and others},
  year   = {2020},
  howpublished = {\url{https://www.wri.org/research/guidance-calculating-water-use-embedded-purchased-electricity}},
}

@misc{aws_sustainability,
  title  = {{AWS} Cloud Sustainability},
  author = {{Amazon Web Services}},
  year   = {2024},
  howpublished = {\url{https://aws.amazon.com/sustainability/}},
}

@misc{google_datacenter_efficiency,
  title  = {Data Centers: Efficiency},
  author = {{Google}},
  year   = {2024},
  howpublished = {\url{https://datacenters.google/efficiency/}},
}

@misc{azure_datacenter_sustainability,
  title  = {Datacenters: Sustainability},
  author = {{Microsoft}},
  year   = {2024},
  howpublished = {\url{https://datacenters.microsoft.com/sustainability/}},
}

@misc{eia_open_data,
  title  = {Open Data},
  author = {{U.S. Energy Information Administration}},
  year   = {2024},
  howpublished = {\url{https://www.eia.gov/opendata/}},
}

@misc{epa_egrid,
  title  = {{eGRID}: Emissions \& Generation Resource Integrated Database},
  author = {{U.S. Environmental Protection Agency}},
  year   = {2024},
  howpublished = {\url{https://www.epa.gov/egrid}},
}

@misc{entsoe_transparency,
  title  = {{ENTSO-E} Transparency Platform},
  author = {{European Network of Transmission System Operators for Electricity}},
  year   = {2024},
  howpublished = {\url{https://transparency.entsoe.eu/}},
}

@misc{uk_carbon_intensity_api,
  title  = {Carbon Intensity {API}},
  author = {{National Energy System Operator}},
  year   = {2024},
  howpublished = {\url{https://carbonintensity.org.uk/}},
}

@misc{rte_eco2mix,
  title  = {{\'e}{CO2}mix: Real-Time Electricity Data for {France}},
  author = {{R{\'e}seau de Transport d'{\'E}lectricit{\'e}}},
  year   = {2024},
  howpublished = {\url{https://www.rte-france.com/eco2mix}},
}

@misc{ember_energy,
  title  = {Global Electricity Data},
  author = {{Ember}},
  year   = {2024},
  howpublished = {\url{https://ember-energy.org/}},
}

@incollection{ipcc_ar6_wg3_ch6,
  title     = {Energy Systems},
  author    = {Clarke, Leon and others},
  booktitle = {Climate Change 2022: Mitigation of Climate Change. Contribution of Working Group {III} to the Sixth Assessment Report of the Intergovernmental Panel on Climate Change},
  publisher = {Cambridge University Press},
  year      = {2022},
}

@misc{iea_energy_and_ai_2025,
  title  = {Energy and {AI}},
  author = {{International Energy Agency}},
  year   = {2025},
  howpublished = {\url{https://www.iea.org/reports/energy-and-ai/energy-demand-from-ai}},
}

@misc{iea_japan_electricity,
  title  = {Japan: Electricity},
  author = {{International Energy Agency}},
  year   = {2025},
  howpublished = {\url{https://www.iea.org/countries/japan/electricity}},
}

@misc{iea_emission_factors_2024,
  title  = {Emissions Factors 2024},
  author = {{International Energy Agency}},
  year   = {2024},
  howpublished = {\url{https://www.iea.org/data-and-statistics/data-product/emissions-factors-2024}},
}

@misc{ghg_protocol_scope2,
  title  = {{GHG} {Protocol} {Scope} 2 {Guidance}},
  author = {{World Resources Institute and World Business Council for Sustainable Development}},
  year   = {2015},
  howpublished = {\url{https://ghgprotocol.org/scope-2-guidance}},
}

@misc{ghg_protocol_corporate_standard,
  title  = {The {Greenhouse} {Gas} {Protocol}: A {Corporate} {Accounting} and {Reporting} {Standard} (Revised Edition)},
  author = {{World Resources Institute and World Business Council for Sustainable Development}},
  year   = {2004},
  howpublished = {\url{https://ghgprotocol.org/corporate-standard}},
}

@misc{schneider_electric_scope3_2023,
  title  = {Demystifying data center scope 3 carbon with our findings},
  author = {{Schneider Electric}},
  year   = {2023},
  howpublished = {\url{https://blog.se.com/datacenter/2023/07/11/demystifying-data-center-scope-3-carbon-with-our-findings/}},
  note   = {White Paper 99, Quantifying Data Center Scope 3 GHG Emissions},
}

@article{oviedo2026energy,
  title   = {Energy use of {AI} inference, efficiency pathways, and test-time scaling},
  author  = {Oviedo, Felipe and others},
  journal = {Joule},
  year    = {2026},
  doi     = {10.1016/j.joule.2026.102430},
}

@misc{zhu2026tracelab,
  title  = {{TraceLab}: Characterizing Coding Agent Workloads for {LLM} Serving},
  author = {Zhu, Kan and others},
  year   = {2026},
  howpublished = {arXiv:2606.30560},
}

\appendix


\section{Benchmarked open-weights models per configuration}
\label{app:model_tables}

Tables~\ref{tab:models_config_h100_fp8},
\ref{tab:models_config_b200_fp8}
and~\ref{tab:models_config_b200_nvfp4} list every benchmarked
open-weights model with its
provider and active/total parameter counts, used to train the
regression models, split up by hardware/quantisation configuration,
sorted smallest to largest by total parameter count.

\begin{table}[htbp]
  \centering
  \caption{Benchmarked open-weights models on H100, FP8. Provider, active and total parameter counts are listed in ascending $N_{\mathrm{total}}$ order.}
  \label{tab:models_config_h100_fp8}
  \footnotesize
  \begin{tabular}{l l r r}
    \toprule
    \textbf{Model} & \textbf{Provider} & \textbf{$N_{\mathrm{active}}$ (B)} & \textbf{$N_{\mathrm{total}}$ (B)} \\
    \midrule
    Qwen3-1.7B-FP8 & Alibaba & 1.7 & 1.7 \\
    Qwen3-4B-Instruct-2507-FP8 & Alibaba & 4.0 & 4.0 \\
    DeepSeek-R1-Distill-Qwen-7B-FP8 & DeepSeek & 7.0 & 7.0 \\
    Mistral-7B-Instruct-v0.3-FP8 & Mistral AI & 7.0 & 7.0 \\
    OpenHermes-2.5-Mistral-7B-FP8-Dynamic & Nous Research & 7.0 & 7.0 \\
    Llama-3-8B-Instruct-FP8 & Meta & 8.0 & 8.0 \\
    Qwen3-8B-FP8 & Alibaba & 8.0 & 8.0 \\
    Qwen3-30B-A3B-FP8 & Alibaba & 3.3 & 30.0 \\
    Qwen3-30B-A3B-Instruct-2507-FP8 & Alibaba & 3.3 & 30.0 \\
    NextCoder-32B-FP8 & Microsoft & 32.0 & 32.0 \\
    Granite-34B-Code-FP8 & IBM & 34.0 & 34.0 \\
    Mixtral-8x7B-Instruct-FP8 & Mistral AI & 12.9 & 46.7 \\
    Hermes-3-Llama-3.1-70B-FP8 & Nous Research & 70.0 & 70.0 \\
    Llama-3.1-70B-Instruct-FP8 & Meta & 70.0 & 70.0 \\
    Llama-3.3-70B-Instruct-FP8 & Meta & 70.0 & 70.0 \\
    Qwen2.5-72B-Instruct-FP8 & Alibaba & 72.0 & 72.0 \\
    Qwen1.5-110B-Chat-GPTQ-Int8 & Alibaba & 110.0 & 110.0 \\
    Mixtral-8x22B-Instruct-FP8 & Mistral AI & 39.0 & 141.0 \\
    DeepSeek-Coder-V2-Instruct-FP8 & DeepSeek & 21.0 & 236.0 \\
    DeepSeek-V2.5-1210-FP8 & DeepSeek & 21.0 & 236.0 \\
    Llama-4-Maverick-17B-128E-Instruct-FP8 & Meta & 17.0 & 400.0 \\
    Llama-3.1-405B-Instruct-FP8 & Meta & 405.0 & 405.0 \\
    \bottomrule
  \end{tabular}
\end{table}

\begin{table}[htbp]
  \centering
  \caption{Benchmarked open-weights models on B200, FP8.}
  \label{tab:models_config_b200_fp8}
  \footnotesize
  \begin{tabular}{l l r r}
    \toprule
    \textbf{Model} & \textbf{Provider} & \textbf{$N_{\mathrm{active}}$ (B)} & \textbf{$N_{\mathrm{total}}$ (B)} \\
    \midrule
    DBRX-Instruct-FP8-KV & Databricks & 36.0 & 132.0 \\
    Mixtral-8x22B-Instruct-FP8 & Mistral AI & 39.0 & 141.0 \\
    GLM-4.7-FP8 & Z.ai & 32.0 & 358.0 \\
    Hermes-4-405B-FP8 & Nous Research & 406.0 & 406.0 \\
    Qwen3-Coder-480B-A35B-Instruct-FP8 & Alibaba & 35.0 & 480.0 \\
    Ling-1T-FP8 & Ling-1t.ai & 50.0 & 1000.0 \\
    \bottomrule
  \end{tabular}
\end{table}

\begin{table}[htbp]
  \centering
  \caption{Benchmarked open-weights models on B200, NVFP4.}
  \label{tab:models_config_b200_nvfp4}
  \footnotesize
  \begin{tabular}{l l r r}
    \toprule
    \textbf{Model} & \textbf{Provider} & \textbf{$N_{\mathrm{active}}$ (B)} & \textbf{$N_{\mathrm{total}}$ (B)} \\
    \midrule
    Phi-4-reasoning-plus-NVFP4 & Microsoft & 14.0 & 14.0 \\
    Qwen3-30B-A3B-NVFP4 & Alibaba & 3.3 & 30.0 \\
    Qwen3-235B-A22B-NVFP4 & Alibaba & 22.0 & 235.0 \\
    DeepSeek-V3-0324-NVFP4 & DeepSeek & 37.0 & 671.0 \\
    Kimi-K2-Thinking-NVFP4 & Kimi AI & 32.0 & 1000.0 \\
    \bottomrule
  \end{tabular}
\end{table}

\end{document}